\documentclass[fleqn,usenatbib]{mnras}

\usepackage{newtxtext,newtxmath}

\usepackage[T1]{fontenc}

\DeclareRobustCommand{\VAN}[3]{#2}
\let\VANthebibliography\thebibliography
\def\thebibliography{\DeclareRobustCommand{\VAN}[3]{##3}\VANthebibliography}

\usepackage{graphicx}	
\usepackage{amsmath}	
\usepackage{siunitx}

\title[L1688's 3D magnetic field and dynamics]{Characterizing three-dimensional magnetic field, turbulence, and self-gravity in the star-forming region L1688}

\author[Hu \& Lazarian]{
Yue Hu$^{1,2}$\thanks{E-mail: yue.hu@wisc.edu}, A. Lazarian$^{2,3}$\thanks{E-mail: alazarian@facstaff.wisc.edu}
\\
$^{1}$Department of Physics, University of Wisconsin-Madison, Madison, WI, 53706, USA\\
$^{2}$Department of Astronomy, University of Wisconsin-Madison, Madison, WI, 53706, USA\\
$^{3}$Centro de Investigación en Astronomía, Universidad Bernardo O’Higgins, Santiago, General Gana 1760, 8370993,
Chile\\
}

\date{Accepted XXX. Received YYY; in original form ZZZ}

\pubyear{2022}

\begin{document}
\label{firstpage}
\pagerange{\pageref{firstpage}--\pageref{lastpage}}
\maketitle

\begin{abstract}
Interaction of three-dimensional magnetic fields, turbulence, and self-gravity in the molecular cloud is crucial in understanding star formation but has not been addressed so far. In this work, we target the low-mass star-forming region L1688 and use the spectral emissions of $^{12}$CO, $^{13}$CO, C$^{18}$O, and H I, as well as polarized dust emissions. To obtain the 3D direction of the magnetic field, we employ the novel polarization fraction analysis. In combining with the plane-of-the-sky (POS) magnetic field strength derived from the Davis–Chandrasekhar-Fermi (DCF) method and the new Differential Measure Analysis (DMA) technique, we present the first measurement of L1688's three-dimensional magnetic field, including its orientation and strength. We find that L1688's magnetic field has two statistically different inclination angles. The low-intensity tail has an inclination angle $\approx55^\circ$ on average, while that of the central dense clump is $\approx30^\circ$. We find the global mean value of total magnetic field strength is $B_{\rm tot}\approx$~\SI{135}{\micro G} from DCF and $B_{\rm tot}\approx$~\SI{75}{\micro G} from DMA. We use the velocity gradient technique (VGT) to separate the magnetic fields' POS orientation associated with L1688 and its foreground/background. The magnetic fields' orientations are statistically coherent. The probability density function of H$_2$ column density and VGT reveal that L1688 is potentially undergoing gravitational contraction at large scale $\approx1.0$~pc and gravitational collapse at small scale $\approx0.2$~pc. The gravitational contraction mainly along the magnetic field resulting in an approximate power-law relation $B_{\rm tot}\propto n_{\rm H}^{1/2}$ when volume density $n_{\rm H}$ is less than approximately $6.0\times10^3$ cm$^{-3}$.
\end{abstract}




\section{Introduction}

Magnetic field, turbulence, and self-gravity play vital roles in the evolution of molecular cloud \citep{1981MNRAS.194..809L,Crutcher04,2011MNRAS.416.1436B,2012ARA&A..50...29C,2020ApJ...905..129H,HLS21,Alina20}. On cloud scales, turbulence and magnetic field can provide global support against gravitational collapse and gravitational contraction \citep{1993ApJ...419L..29E,1995MNRAS.277..377P,2000ApJ...535..887K,Crutcher04,Crutcher12,2012ApJ...757..154L,2012ApJ...761..156F,2019MNRAS.490.3061V}. However, turbulence additionally can produce compression and local density enhancements serving as seeds of star formation \citep{MK04,MO07,2012ApJ...761..156F}. In view of the importance, intense efforts to access these three factors have been made.

Typically the significance of turbulence can be measured by the emission line's width due to the Doppler shift effect. Such measurement already revealed the supersonic nature of molecular cloud \citep{2010ApJS..191..232B,2011PASJ...63..105S,2015ApJ...801...25L,2016ApJ...832..143F,HLS21}. Measuring the magnetic field is even more challenging. The plane-of-the-sky (POS) magnetic field orientation can be traced by polarized dust emission based on the property that radiative torques (RATs) align dust grains with their semi-major axis perpendicularly to the magnetic field \citep{2007MNRAS.378..910L,2015ARA&A..53..501A}. The line-of-sight (LOS) magnetic field strength is traced by Zeeman splitting \citep{Crutcher04,2012ARA&A..50...29C} or Faraday rotation \citep{2018A&A...614A.100T}. However, because these measurements alone probe different regions of the multi-phase interstellar medium (ISM) or are sensitive to distinct density ranges, they cannot be easily combined to yield the full 3D magnetic field vector, including both 3D direction and total strength. 

Due to this limitation, observational studies of turbulence, magnetic field, and self-gravity's interaction still stay in two-dimension, i.e. either the POS or the LOS component. For instance, the Davis–Chandrasekhar–Fermi (DCF) method \citep{1951PhRv...81..890D,1953ApJ...118..113C} is widely used to estimate the POS magnetic field strength \citep{HLS21,2021ApJ...907...88P,2021ApJ...913...85H,2021MNRAS.tmp.3119L,2022ApJ...929...27H,2022arXiv220512084T}. The limitations of DCF were discussed extensively in literature \citep{2021A&A...656A.118S,2022arXiv220409731L,2022arXiv220509134C,2022ApJ...925...30L}, and an alternative Differential Alignment Measure (DMA; \citealt{2022arXiv220409731L}) technique was introduced. In this paper, we use both techniques and compare their results.  

To access the 3D magnetic field, we employ the recently proposed dust polarization fraction analysis \citep{2022arXiv220309745H}. This method uses the fact that the linear polarization fraction is sensitive to the efficiency with which grains are aligned with respect to the magnetic field, the POS magnetic field's degree of disorder, and the 3D magnetic field's inclination angle with respect to the LOS. Under the assumption of homogeneous dust grains' properties, the observed polarization fraction in a strongly magnetized region, i.e., with an ordered POS magnetic field angle, is only determined by the inclination angle \citep{2019MNRAS.485.3499C,2022arXiv220309745H}. The inclination angle, therefore, can be calculated from the corresponding polarization fraction based on their correlation given in \cite{2022arXiv220309745H}. This technique is termed Polarization Fraction Analysis (PFA). Combining with DCF or DMA, this technique opens a new way to characterize the 3D magnetic field, including both POS and LOS components' orientation and strength. Especially, both the POS and the LOS components are inferred from the same polarization data set, the two components are not fully independent. PFA ensures the measured LOS and POS magnetic fields are from the same ISM phase and spatial region.

This work aims at providing the first observational analysis of 3D magnetic fields in nearby molecular clouds. We target the star-forming region L1688, which is one of the closest star-forming clouds located at a distance of approximately 139~pc \citep{2008hsf2.book..351W,2018ApJ...869L..33O}. It has a large and varied population of young stellar objects, serving as an excellent object for studying low-mass star formation \citep{2015ApJS..220...11D}. Although L1688 has been investigated over a wide wavelength range providing intensive data sets, the interaction of 3D magnetic field, turbulence, and self-gravity there has never been studied. In the paper below, we use the Planck \citep{2020AA...641A..11P}, HAWC+ \citep{2019ApJ...882..113S} polarized dust emission, and the Velocity Gradient Technique (VGT; \citealt{GL17,YL17a,LY18a,HYL18}) to trace the POS magnetic field. PFA and DCF (or DMA) output the inclination angle and POS magnetic field strength, respectively. In addition, we assess the significance of turbulence from $^{12}$CO (1-0) and $^{13}$CO (1-0) emission lines obtained from the COMPLETE survey  \citep{2006AJ....131.2921R}. The C$^{18}$O (3-2) observed with the APEX telescope provides information on a zoom-in region around Oph A \citep{2010AA...510A..98L}. As for the role of self-gravity, we use the probability density function (PDF; \citealt{2011MNRAS.416.1436B,2012ApJ...750...13C,2018ApJ...863..118B,2019MNRAS.482.5233K,2021MNRAS.502.1768H}) of Herschel's column density \citep{2020A&A...638A..74L} as well as VGT to identify the gravity-dominated region. 

This paper is organized as follows. In \S~\ref{sec:data}, we briefly describe the observational data. In \S~\ref{sec:method}, we introduce the statistic tools used in this work. In \S~\ref{sec:results}, we present our results of the identified self-gravitating region, the 3D magnetic field's orientation and strength, and the comparison of self-gravity, turbulence, and magnetic field's significance in L1688. We discuss the dynamics of L1688 and potential uncertainty in our analysis in \S~\ref{sec:dis} and give a summary in \S~\ref{sec:con}.

\section{Observational data}
\label{sec:data}
\subsection{$^{12}$CO, $^{13}$CO, and C$^{18}$O emission lines}
In this work, we adopt the $^{12}$CO (1-0) and $^{13}$CO (1-0) emission lines over the L1688 cloud provided by the COMPLETE survey \citep{2006AJ....131.2921R}. The observation was performed with the 14m Five College Radio Astronomy Observatory (FCRAO) telescope. Each line was measured with a total bandwidth of 25 MHz and 1024 channels, yielding an effective velocity resolution of 0.07 km/s. The Half Power Beam Width (HPBW) of the observation is $\approx46$" for $^{12}$CO and $\approx44$" for $^{13}$CO. The final data cube, however, is convolved onto a regular $23$" per pixel resolution. The RMS noise level per channel is $\approx0.98$ K for $^{12}$CO and $\approx0.33$ K for $^{13}$CO in unit of antenna temperature $T_{\rm A}^*$. The radial velocity of the cloud's bulk motion ranges from 0 to 7 km/s. We select the emissions within this velocity range for our analysis.

The C$^{18}$O (3-2) emission was observed with the 12m APEX telescope \citep{2010AA...510A..98L}. The observation at 329 GHz has HPBW $\approx19$" and is sampled according to the Nyquist criterion onto 2.5" per pixel. The data cube achieves a velocity resolution $\approx0.11$ km/s with a RMS noise $\approx0.05$ K in unit of $T_{\rm A}^*$. 

\subsection{H I emission line and $\rm H_2$ column density}
Cube of 21 cm H I emission were observed with the 100m NRAO Green Bank Telescope \citep{2003ApJ...585..823L}. The data has beam resolution HPBW $\approx9'$ and is regridded to $4'$ per pixel. The final data have a spectral resolution of 0.32 km/s and the corresponding RMS noise level is $\approx0.15$ K. 

The H$_2$ column density data with beam resolution $\approx18.2$" is obtained from the Herschel Gould Belt Survey \citep{2010AA...518L.102A}. The data of the Ophiuchus molecular cloud was presented in \cite{2020A&A...638A..74L}.

\subsection{Polarized dust emission}
The POS magnetic field orientation was inferred from the Planck 353 GHz and HAWC+ polarized dust thermal emission. In this work, we use the Planck 3rd Public Data Release (DR3) 2018 of High Frequency Instrument \citep{2020A&A...641A...3P}\footnote{Based on observations obtained with Planck (\url{http://www.esa.int/Planck}), an ESA science mission with instruments and contributions directly funded by ESA Member States, NASA, and Canada.} at 353 GHz (HPBW$\approx5'$), HAWC+ band A (\SI{53}{\micro\meter}; HPBW$\approx4.8$"), and HAWC+ band C (\SI{89}{\micro\meter}; HPBW$\approx7.8$"). The polarization angle $\phi$ is defined through Stokes parameters $I$, $Q$, and $U$:
\begin{equation}
\begin{aligned}
   \phi&=\frac{1}{2}\arctan(-U,Q),\\
\end{aligned}
\end{equation}
where $-U$ converts the angle from the HEALPix convention to the IAU convention and the two-argument function $\arctan$ is used to account for the $\pi$ periodicity. The magnetic field angle is inferred from $\phi_B=\phi+\pi/2$.

To correct the observed polarization fraction $p$ for the bias, we adopt the commonly used debiasing method \citep{1974ApJ...194..249W,2019ApJ...880...27P}, under which the debiased $p$ is given by: 
\begin{equation}
\begin{aligned}
   p&=\frac{\sqrt{Q^2+U^2-\frac{1}{2}(\delta Q^2+\delta U^2)}}{I},
\end{aligned}
\end{equation}
where $\delta Q$ and $\delta U$ are the uncertainties in Stokes $Q$ and $U$, respectively. The uncertainties of polarization fraction and polarization angle are presented in Figs.~\ref{fig:sigma_p} and \ref{fig:sigma_phi}.

As for HAWC+, the debaised polarization fraction and corresponding uncertainty maps are directly available in SOFIA data archive (see \citealt{2019ApJ...882..113S}). We select only pixels with $p/\delta p>3$ and $I/\delta I>100$. A summary of the data sets is available in Tab.~1.

\begin{table*}
	\centering
	\label{tab.1}
	\begin{tabular}{| c | c | c | c | c |}
		\hline
		Observation & Frequency or wavelength & Beam resolution & Velocity resolution & Reference \\\hline\hline
		H I & 21 cm & 9$'$ &  0.32 km/s & \cite{2003ApJ...585..823L}  \\
		$^{12}$CO (1-0) & 115.271 GHz & 46" & 0.07 km/s & \cite{2006AJ....131.2921R} \\
		$^{13}$CO (1-0) & 110.201 GHz & 44" & 0.07 km/s & \cite{2006AJ....131.2921R} \\
		C$^{18}$O (3-2) & 329 GHz & 19" & 0.11 km/s & \cite{2010AA...510A..98L} \\
		H$_2$ & - & 18.2" & - & \cite{2010AA...518L.102A} \\
		Planck polarization & 353 GHz & 5$'$ & -  & \cite{2020AA...641A..11P} \\
		HAWC+ polarization & \SI{154}{\micro\meter} \& \SI{89}{\micro\meter} \& \SI{53}{\micro\meter}  & 7.8" \& 4.8" & - & \cite{2019ApJ...882..113S} \\\hline
	\end{tabular}
	\caption{Information of observational data sets used in this work.}
\end{table*}

\section{Methodology}
\label{sec:method}
\subsection{Probability density function of H$_2$ column density}
The PDF of H$_2$ column density is widely used to study turbulence and self-gravity in the ISM. The PDF follows a hybrid of log-normal distribution $P_{\rm LN}(s)$ in diffuse turbulence-dominated region and power-law distribution $P_{\rm PL}(s)$ in gravity-dominated region \citep{2011MNRAS.416.1436B,2012ApJ...750...13C,2018ApJ...863..118B,2019MNRAS.482.5233K,2021MNRAS.502.1768H}:
\begin{equation}
\begin{aligned}
    P_{\rm LN}(s)&\propto\frac{1}{\sqrt{2\pi\sigma_s^2}}e^{-\frac{(s-s_0)^2}{2\sigma_s^2}},s<S_t,\\
    P_{\rm PL}(s)&\propto e^{a s},s>S_t,\\
\end{aligned}
\end{equation}
where $s= \log(N_{\rm H_2}/\overline{N_{\rm H_2}})$ is the logarithm of the column density $N_{\rm H_2}$ normalized by its mean value $\overline{N_{\rm H_2}}$. $S_t$ is transitional density between the $P_{\rm PL}$ and $P_{\rm LN}$. $s_0$ is the mean logarithmic density, $\sigma_s$ represents the standard deviation, and $a$ stands for the power-law distribution's slope.
The value of $a$ changes in roughly the cloud mean free-fall time from steep {$\approx -3$ to shallow $\approx -1.5$.} A shallow slope indicates a high star formation rate \citep{2018ApJ...863..118B}.

\subsection{Velocity gradient technique}
VGT is a novel approach to tracing the POS magnetic field in molecular clouds. VGT relies on the theories of MHD turbulence and turbulent reconnection \citep{GS95,LV99}. It used the fact that the magnetized and turbulent eddies are anisotropic, i.e., elongating along the direction of magnetic field. The anisotropy means maximum velocity fluctuation appears in the direction perpendicular to the magnetic field. Consequently, the gradient of velocity fluctuation is also perpendicular to the magnetic field. However, in the case that gravitational collapse dominates over MHD turbulence, the velocity gradient's orientation is determined by the infall acceleration being parallel to the magnetic field \citep{HLY20,HLS21}. The properties of velocity gradients were first theoretically predicted and then confirmed observationally. 

Obtaining velocity information from observations is not trivial. One of the possible ways is using thin velocity channels, which are dominated by the effect of velocity caustics \citep{LP00,2017MNRAS.464.3617K}. The effect means that real density structures with different velocities along the LOS are sampled into different velocity channels. The observed intensity structures in a velocity channel are distorted due to turbulence. When the velocity channel is narrow enough (i.e., the channel width is smaller than the velocity dispersion), the distortion is significant, and the observed intensity fluctuation is dominated by velocity fluctuation rather than density fluctuation \citep{LP00}. One can, therefore, use the thin velocity channels to get velocity fluctuations and calculate the velocity gradient. Further improvements are possible on the basis of the general theory of Position-Position-Velocity (PPV) statistics developed in \cite{LP00}. 

We adopt the VGT recipe used in \cite{2022MNRAS.511..829H} and briefly describe it here. (i) each thin channel map was convolved with a 3×3 Sobel Kernel to create a raw gradient map (pixels are blanked out if their intensity is less than three times the RMS noise level); (ii) the mean gradient angle at each pixel is statistically calculated from the sub-block average method, in which the histogram of raw gradients' orientation within a rectangle sub-block (size = $20\times20$ pixels, which is a numerically and empirically optimal value; \citealt{LY18a,HLS21}.) is fitted with a Gaussian distribution to find the most probable orientation; (iii) the cosine and sine values of each sub-block averaged gradient map is weighted with the corresponding channel's intensity and integrated along the LOS to construct pseudo-Stokes parameter $Q_{\rm g}$ and $U_{\rm g}$. (iv) the POS magnetic field orientation inferred from VGT is then $\phi_B^{\rm VGT}=\frac{1}{2}\arctan(U_{\rm g},Q_{\rm g})+\pi/2$;

The alignment between VGT measurement and polarization measurement is quantified by the Alignment Measure (AM; \citealt{GL17}), defined as:
\begin{equation}
        \begin{aligned}
           {\rm AM}  =  2(\cos^2\theta_r-\frac{1}{2}),
        \end{aligned}
\end{equation}
where $\theta_r$ is the relative angle between the two angles. AM is a relative scale ranging from -1 to 1, with AM = 1 indicating that two angles are parallel and AM = -1 denoting that the two are orthogonal.  

\begin{figure}
	\includegraphics[width=1.0\linewidth]{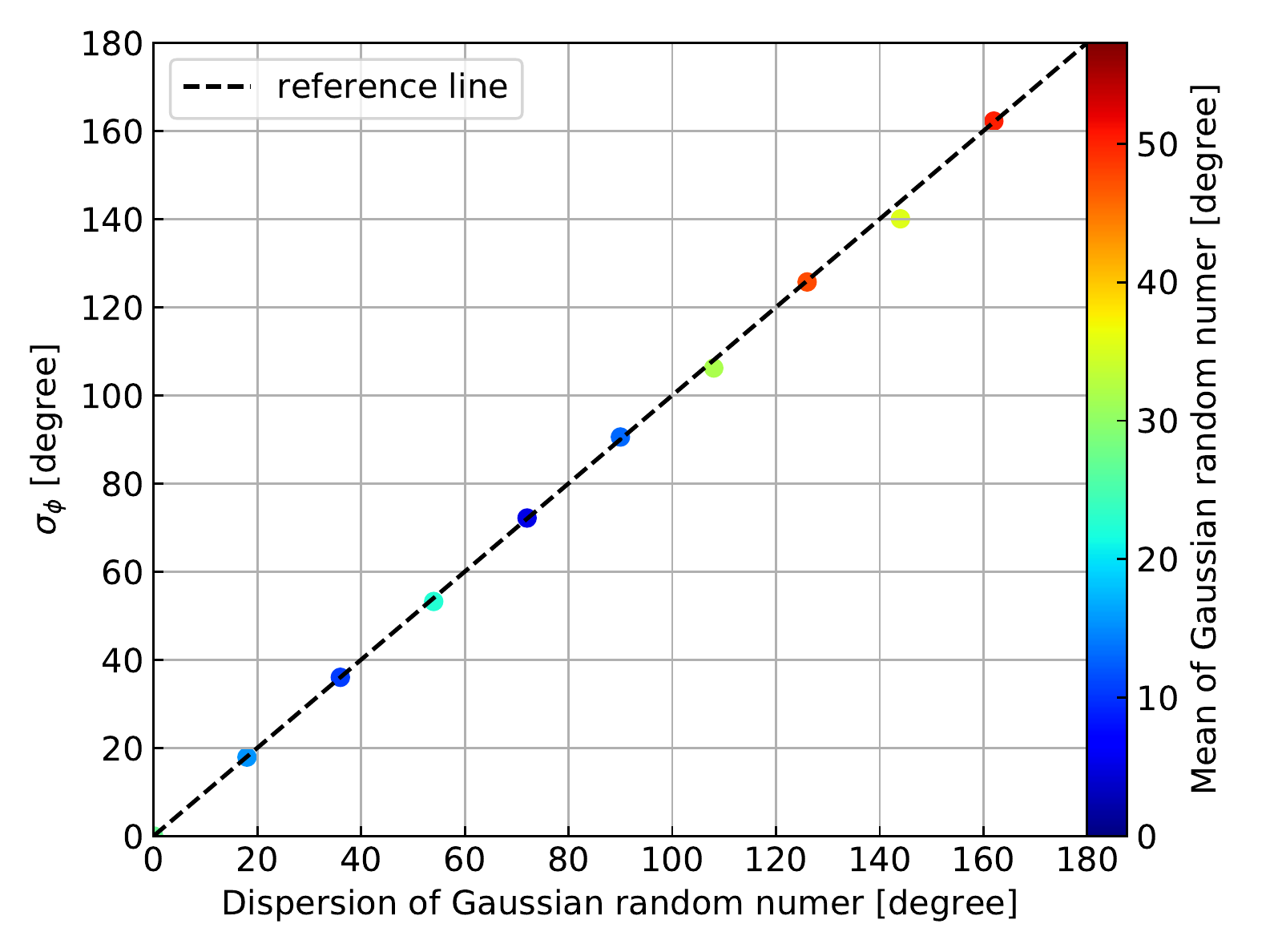}
    \caption{Comparison of the angle dispersion estimated from Eq.~\ref{eq.sigma_phi} and the real dispersion generated from Gaussian random numbers. Colors represent the mean (randomly generated) of the Gaussian random numbers.}
    \label{fig:Gdis}
\end{figure}

\subsection{Measuring magnetic field}
\subsubsection{Davis–Chandrasekhar-Fermi (DCF) method}
{\bf Traditional DCF:}\\
The POS magnetic field strength can be obtained via the DCF method \citep{1951PhRv...81..890D,1953ApJ...118..113C}. Assuming the variations of the magnetic field direction arise from isotropic incompressible turbulence and ordered variation in the magnetic field is insignificant, the magnetic field strength is expressed as:
\begin{equation}
     B=f\sqrt{4\pi\rho}\sigma_{v}/\sigma_\phi,
\end{equation}
where $f=0.5$ is an empirically chosen correction factor \citep{2001ApJ...546..980O}, $\rho$ is the volume mass density. $\sigma_{v}$ and $\sigma_\phi$ represent 1D turbulent velocity dispersion along the LOS and polarization angle's dispersion on the POS. The uncertainty of polarization angle (given in Fig.~\ref{fig:sigma_phi} with a median value of $\approx14^\circ$) may affect $\sigma_\phi$. In any case, we discuss more about the uncertainties related to the DCF method in \S~\ref{subsec:uncer}.

{\bf \noindent Modification of DCF:}\\
The shortcoming of the traditional DCF is that the dispersion of angles is calculated about the mean, the latter being a poorly defined quantity in observations. To deal with this problem we propose a modified DCF, where we calculate the angle dispersion in a different way. In our study, $\sigma_\phi$ is characterized by angular statistics \citep{fisher1995statistical,Lazarian18}:
\begin{equation}
\label{eq.sigma_phi}
    \sigma_\phi=\sqrt{-\log(\langle\cos\phi\rangle^2+\langle\sin\phi\rangle^2)},
\end{equation}
where $\langle...\rangle$ stands for ensemble average. Note that in the formula, $\phi$ is in the range of $[-\pi,\pi]$, but the polarization angle in observational data is in the range of $[-\pi/2,\pi/2]$. Therefore, we double the polarization angle when processing the observational data and divide $\sigma_\phi$ by two as the final result. Fig.~\ref{fig:Gdis} presents a simple numerical test of Eq.~\ref{eq.sigma_phi}. We generate 10 sets of Gaussian random numbers (1000 numbers per set). Each set is assigned a dispersion and a random mean value. We see Eq.~\ref{eq.sigma_phi} accurately recovers the dispersion. For our purpose of using the DCF method, $\sigma_\phi$ is estimated for each sub-block with size $30\times30$ pixels (pixel resolution: $\approx23''$) covering approximately $3\times3$ independent beams of Planck observation. The uncertainties raised by the choice of sub-block size and other factors are discussed in \S~\ref{subsec:uncer}.

\subsubsection{Differential Measure Approach (DMA)}
One of the limitations of DCF is that it is difficult to obtain the dispersion of the angles and velocities over limited areas (see \S~\ref{sec:dis}). The subtraction of the thermal speed is also subject to errors. Moreover, DCF assumes that the observed fluctuations are purely raised Alfv\'en waves on the POS. The real MHD turbulence, however, is composed of compressible modes \citep{2003MNRAS.345..325C}.  

These and other problems of DCF were addressed by the DMA technique in \citet{2022arXiv220409731L}. In what follows, we employ the simplest, version of the DMA. It was shown in \citet{2022arXiv220409731L} that even this "naive" version of the DMA provides better recovery of the magnetic field values compared to DCF. That simplified version can be presented as:
\begin{equation}
\label{eq.dma}
         B=f'\sqrt{4\pi\rho}\sqrt{D_v(l)/D_\phi(l)},
\end{equation}
where $D_v$ and $D_\phi$ are the second order structure function of velocity centroid $C(\pmb{r})$ and polarization angel at scale $l$, respectively. The adjustment factor $f'$ depends on the composition of turbulence in terms of three basic MHD modes \citep{2003MNRAS.345..325C} and the magnetic field's inclination angle. Here we choose a constant scaling factor $f'=0.5$ for the purpose of comparing DCF and DMA.


The quantities that enter the structure functions above are defined as:
\begin{equation}
\label{eq.D}
    \begin{aligned}
        C(\pmb{r})&=\int T_{\rm mb}(\pmb{r},v)v dv/\int T_{\rm mb}(\pmb{r},v) dv,\\
        D_v(l)&=\langle (C(\pmb{r})-C(\pmb{r}+\pmb{l}))^2\rangle,\\
        D_\phi(l)&=\langle (\phi(\pmb{r})-\phi(\pmb{r}+\pmb{l}))^2\rangle,\\
    \end{aligned}
\end{equation}
where $T_{\rm mb}(\pmb{r},v)$ is $^{12}$CO brightness temperature, $v$ is the LOS velocity, and $\pmb{r}=(x,y)$ is the spatial position on the POS. Note, that $D_v$ and $D_\phi$ are scale-dependent, but their ratio is not and can give the mean magnetic field strength over a sub-block of interests via Eq.~\ref{eq.dma}. The sub-block size is selected as $30\times30$ pixels to keep consistent with the DCF method. 

\subsection{Polarization fraction analysis (PFA)}
To access the angle $\gamma$ that the total magnetic field inclined to the LOS, we use the PFA proposed in \cite{2022arXiv220309745H}. It is well known that the polarization fraction is correlated with dust grains' intrinsic properties, the POS magnetic field's degree of disorder, and the 3D magnetic field's inclination angle \footnote{ Note PFA is measuring the (effective) mean inclination angle of an accumulation of various magnetic fields along the LOS.} \citep{2019MNRAS.485.3499C,2022arXiv220309745H}. Assuming homogeneous conditions for dust grains, in a strongly magnetized region, $M_{\rm A, POS}^2\ll1$, the fraction is mainly determined by the inclination angle. Here $M_{\rm A, POS}$ is the POS Alfv\'en Mach number. Obtaining the distribution of $M_{\rm A, POS}$ is non-trivial in observation\footnote{The distribution of $M_{\rm A, POS}$ can be obtained from the velocity gradient's dispersion \citep{Lazarian18,Hu19a} or the structure function analysis of velocity centroid \citep{2021ApJ...910...88X,2021ApJ...911...37H}.}. However, $M_{\rm A, POS}$ is positively correlated with the dispersion of polarization angle, so we simplify the condition by using $\sigma_\phi\sim M_{\rm A, POS}$ \citep{2008ApJ...679..537F}. Therefore, for such a region, we have \citep{2022arXiv220309745H}:
\begin{equation}
\label{eq.gamma}
    \sin^2\gamma=\frac{p_{\rm off}(1+p_{\rm max})}{p_{\rm max}(1+p_{\rm off})},~~~ \sigma_\phi^2\ll1,
\end{equation}
where $p_{\rm max}$ is the maximum polarization fraction observed in a cloud and $p_{\rm off}$ is the polarization fraction corresponding to the region with $\sigma_\phi^2\ll1$. $\sigma_\phi^2$ is a second-order quantity so its uncertainty is even smaller and have only minimum effect.

The procedure to calculate the $\gamma$ distribution is then: (i) calculating the map of $\sigma_\phi$ and finding the maximum value of $p$ across the full cloud; (ii) dividing the map into a number of sub-blocks and getting the non-zero polarization fraction $p_{\rm off}$ corresponding to minimum $\sigma_\phi$ for each sub-block. To reduce uncertainty, here we take the polarization fraction's average in the five positions satisfying $\sigma_\phi^2\ll1$ as $p_{\rm off}$. (iii) computing $\gamma$ for each sub-block using Eq.~\ref{eq.gamma}. Same as the calculation of $\sigma_\phi$, we set the sub-block size to $30\times30$ pixels to match the polarization dispersion map. 

Note that PFA is different from another method proposed in \cite{2019MNRAS.485.3499C}. There the observed polarization fraction in every pixel, together with $p_{\rm max}$, is used to derive a pixelized $\gamma$ distribution. This method, however, assumes the polarization fraction is determined only by the inclination angle neglecting all magnetic field's fluctuations, which could significantly decrease the polarization fraction in a turbulent molecular cloud.

\begin{figure*}
	\includegraphics[width=1.0\linewidth]{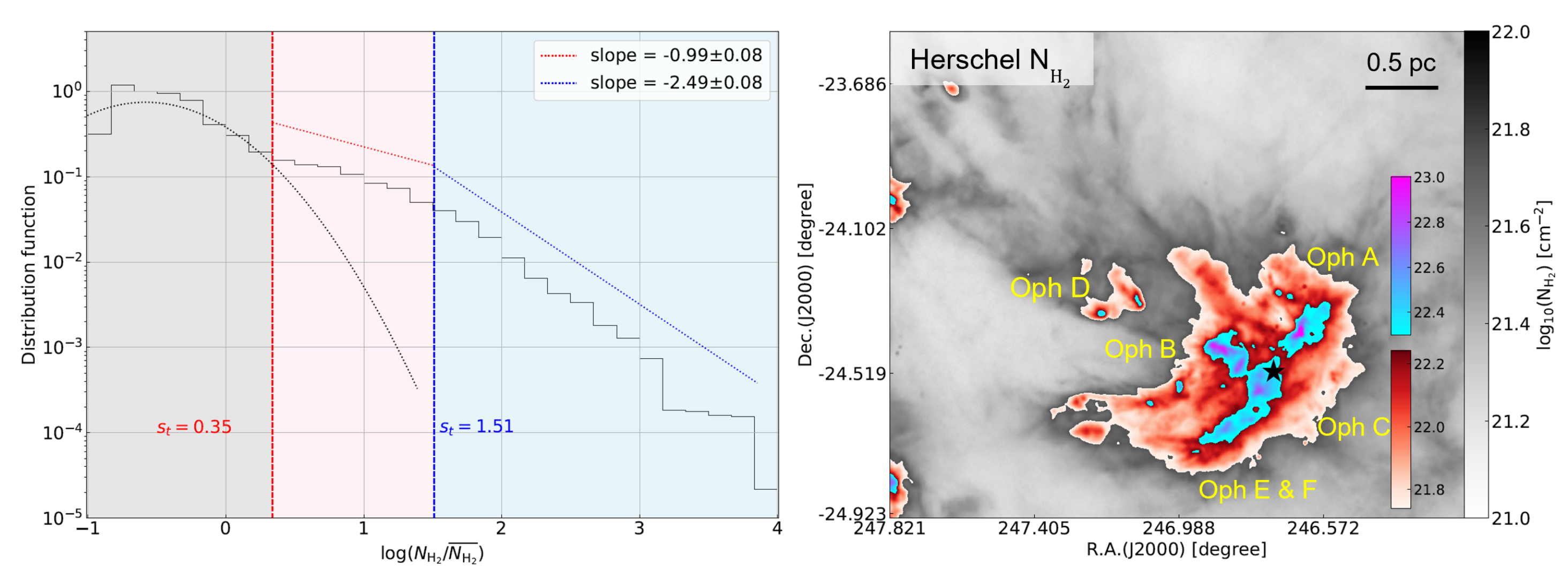}
    \caption{\textbf{Left:} the PDFs of the ${\rm H_2}$ column density. Dotted lines represent the best fitting of log-normal and power-law distributions. $S_t$ is the transition density threshold. \textbf{Right:} Map of ${\rm H_2}$ column density ${\rm N_{H_2}}$ obtained from the Herschel Gould Belt Survey. Red and blue areas indicate the gravitationally collapsing regions identified from the PDFs. We assume the cross point denoted by a star symbol (R.A.$\approx246.65^\circ$, and Dec.$\approx-24.52^\circ$) of Oph A and B's elongation as the cloud center.}
    \label{fig:NH2}
\end{figure*}

\begin{figure}
	\includegraphics[width=1.0\linewidth]{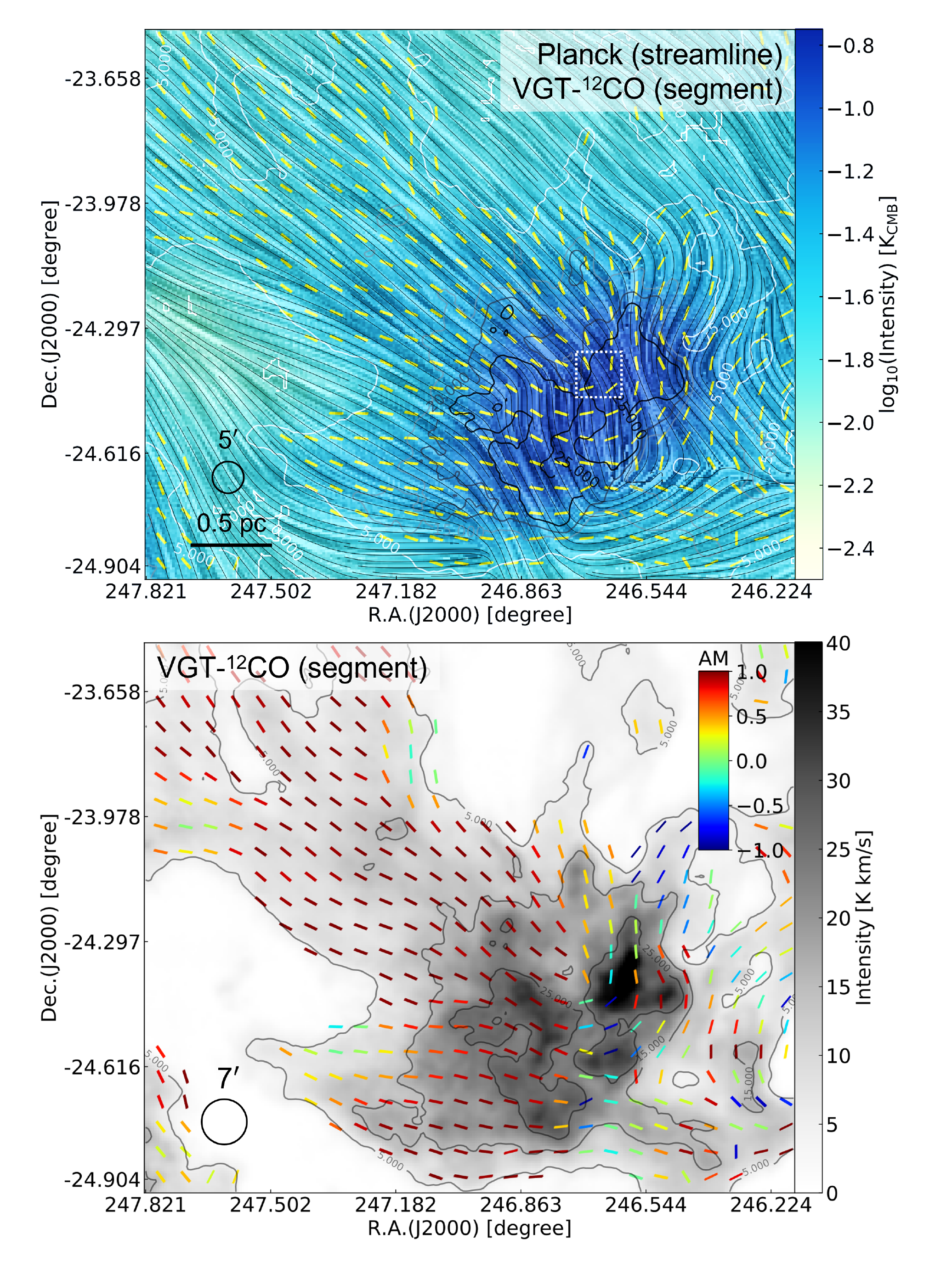}
    \caption{\textbf{Top:} Morphology of magnetic fields revealed by the Planck polarization (streamlines) at 353 GHz and VGT (yellow segment) using $^{12}$CO (J = 1-0) emission. The magnetic field is overlaid with the intensity map of 353 GHz dust emission. The black circle represents the beamwidth of observation. Contours outline the intensity structures of $^{12}$CO starting from 5 K km/s. \textbf{Middle:} Morphology of magnetic fields revealed by VGT (colored segments) using $^{12}$CO emission. Colors on vectors present the AM of VGT and Planck polarization.}
    \label{fig:plank}
\end{figure}

\begin{figure}
	\includegraphics[width=1.0\linewidth]{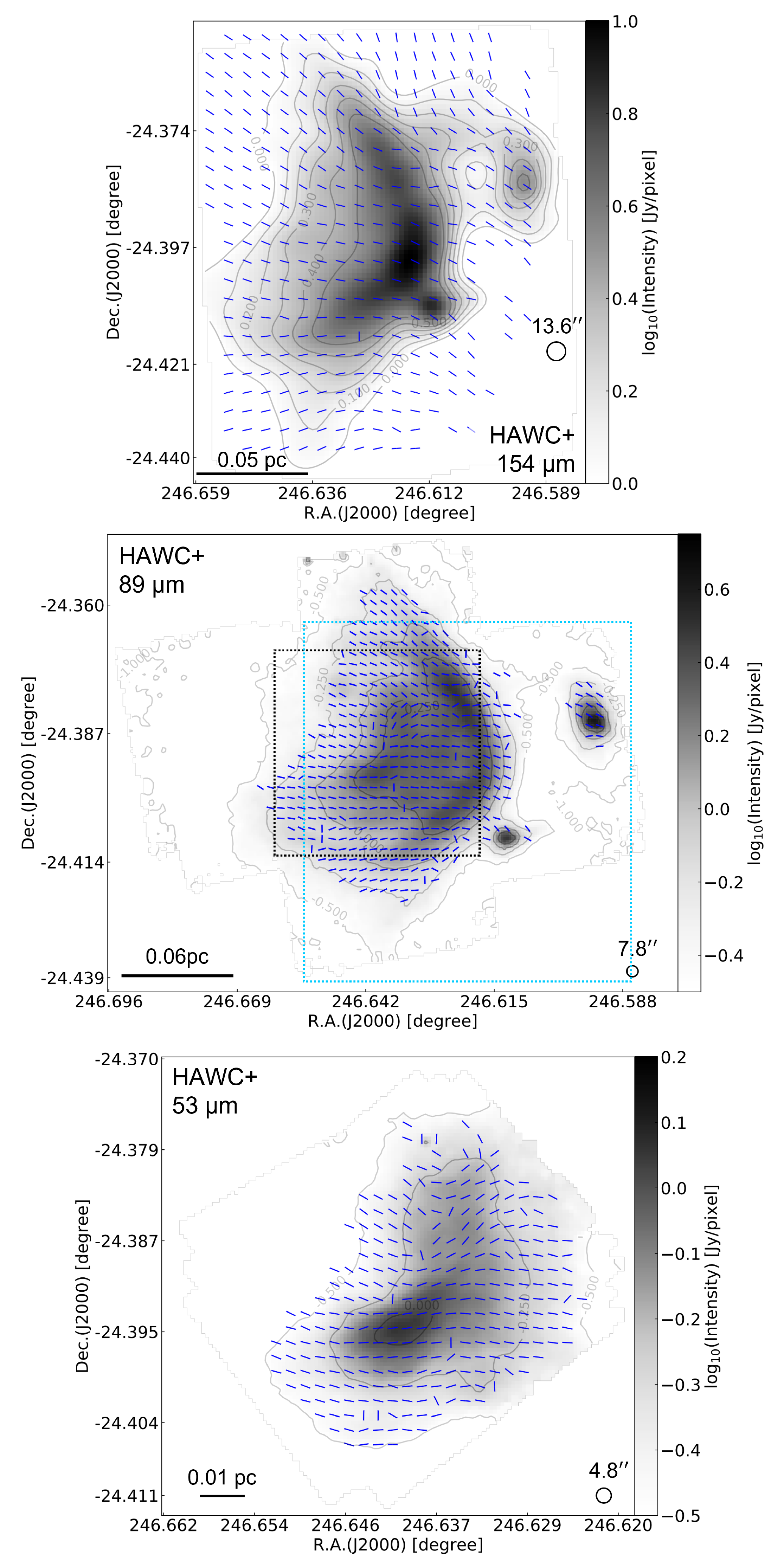}
    \caption{\textbf{Top:} magnetic fields inferred from HAWC+ polarization at \SI{154}{\micro\meter} for a zoom-in region as indicated in the middle panel by a blue dashed box. \textbf{Middle:} magnetic fields inferred from HAWC+ polarization at \SI{89}{\micro\meter} for a zoom-in region as indicated in Fig.~\ref{fig:plank} by a dashed box. \textbf{Bottom:} magnetic fields inferred from HAWC+ polarization at \SI{53}{\micro\meter} for a zoom-in region as indicated in the middle panel by a black dashed box.}
    \label{fig:hawc}
\end{figure}

\begin{figure}
	\includegraphics[width=1.0\linewidth]{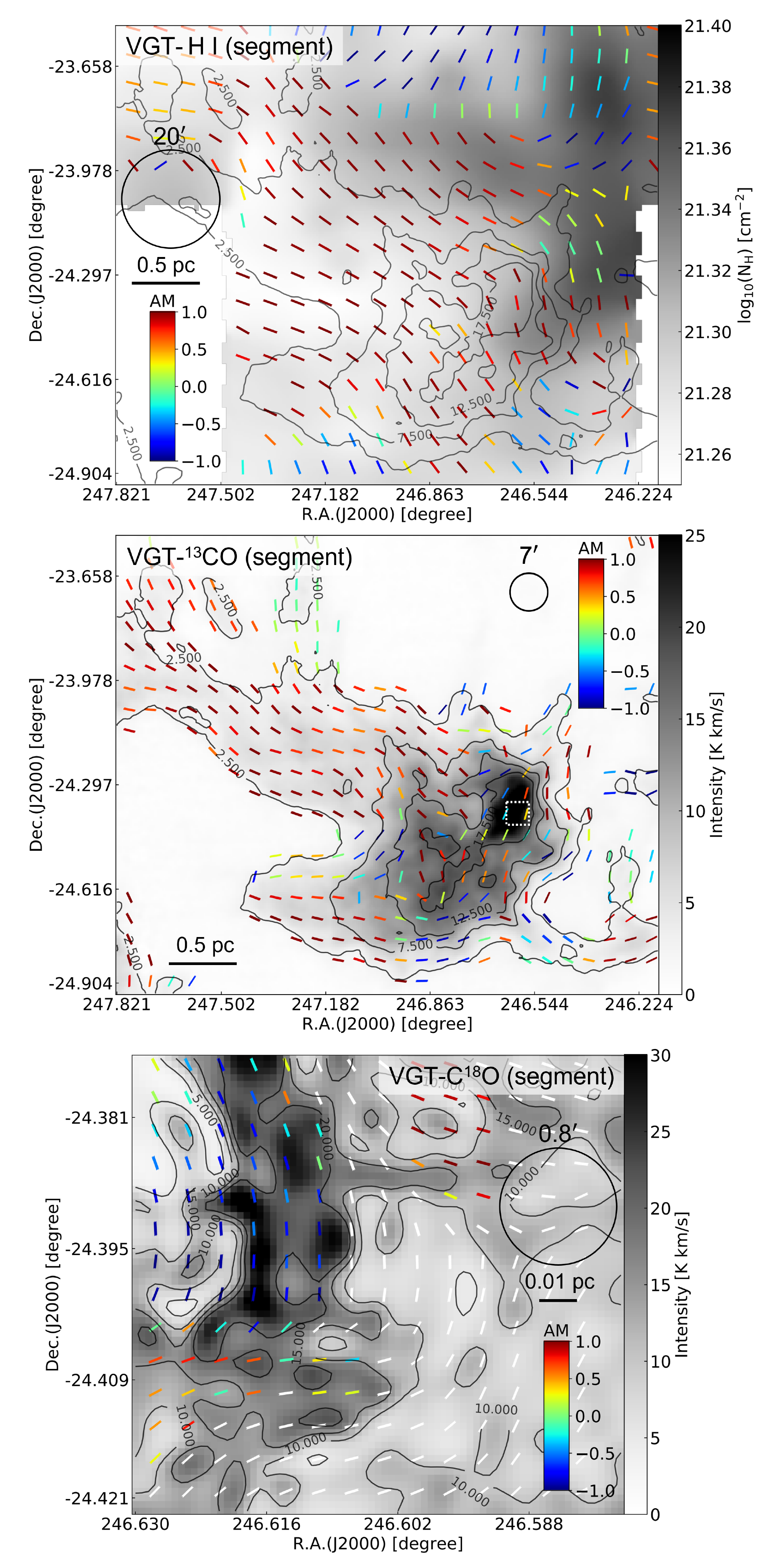}
    \caption{\textbf{Top:} Morphology of magnetic fields revealed by VGT (colored segments) using H I emission. The magnetic field is overlaid with the $\rm N_{H}$ column density map. Colors on vectors present the AM of VGT and Planck polarization. Contours outline the intensity structures of $^{13}$CO starting from 2.5 K km/s. \textbf{Middle:} same as top panel, but using $^{13}$CO emission. \textbf{Bottom:} magnetic fields inferred from VGT (colored segments) using C$^{18}$O emission for a zoom-in region as indicated in the middle panel by a dashed box. Colors, except white, on polarization vectors, present the AM of VGT and HAWC+ polarization at \SI{89}{\micro\meter} polarization. White color means no corresponding polarization measurement.}
    \label{fig:co}
\end{figure}

\begin{figure}
	\includegraphics[width=1.0\linewidth]{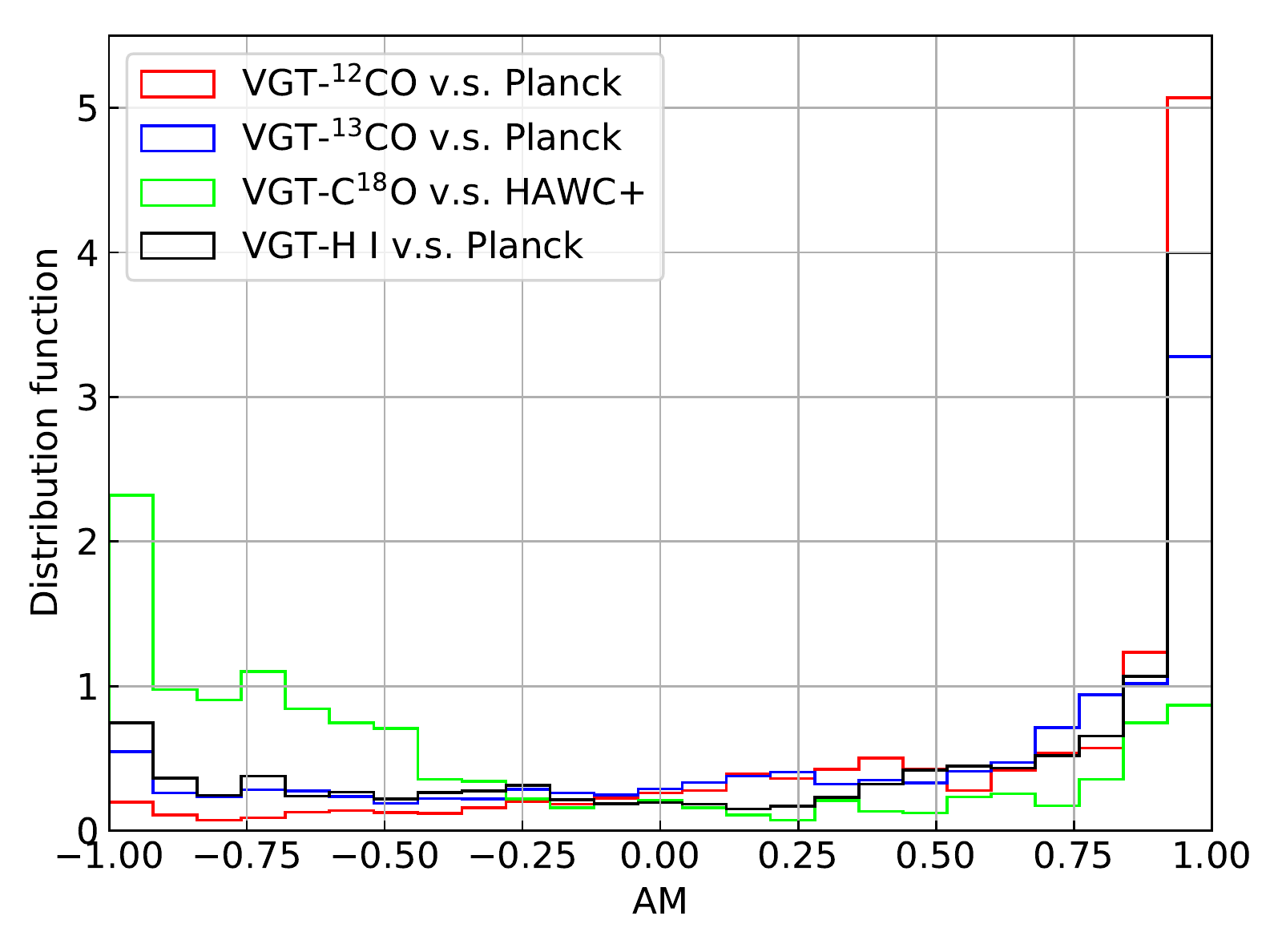}
    \caption{Histogram of the AM calculated from VGT and polarization measurements.}
    \label{fig:AM}
\end{figure}

\begin{figure}
	\includegraphics[width=1.05\linewidth]{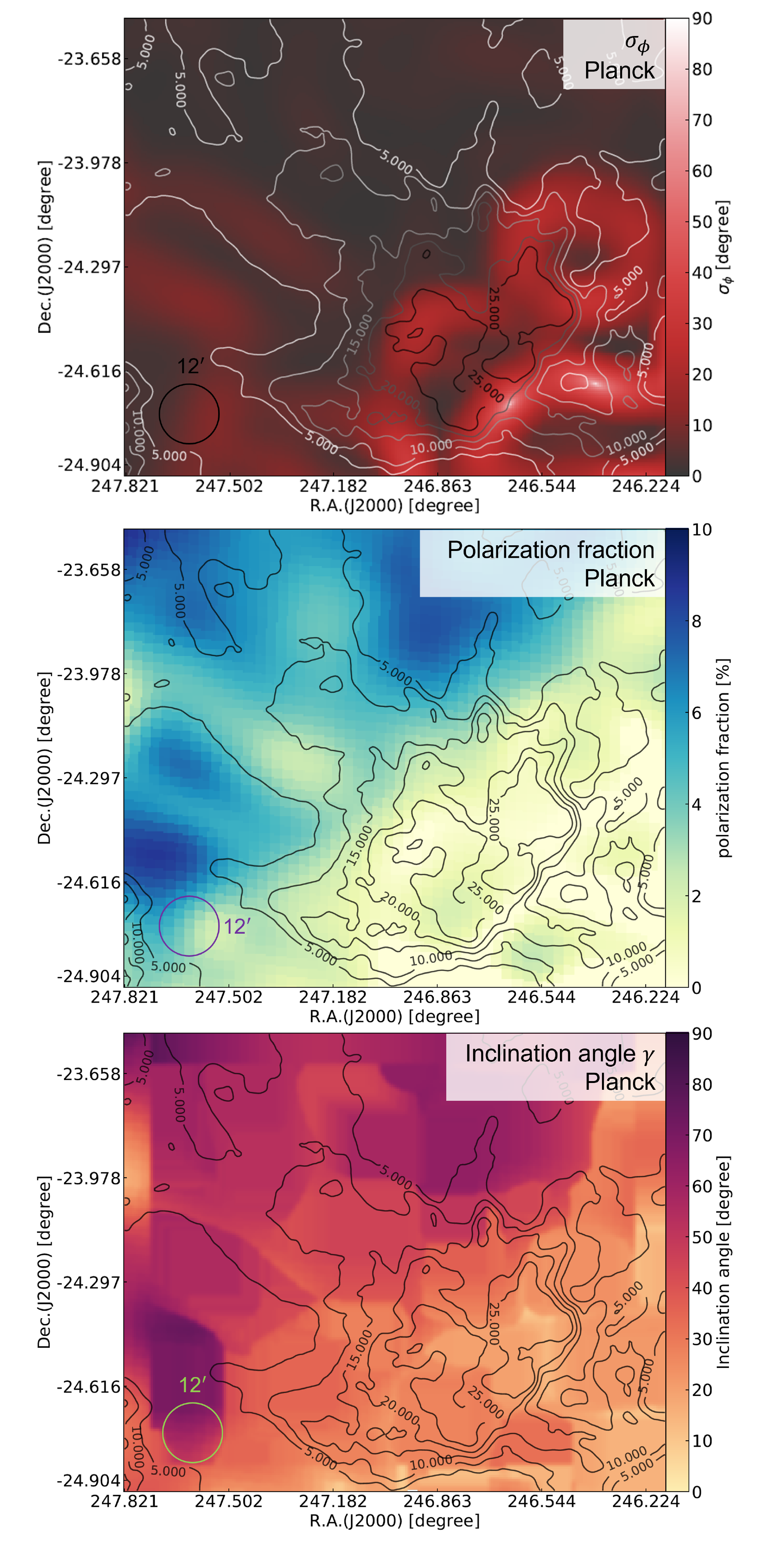}
    \caption{\textbf{Top:} the dispersion of (Planck) polarization angle. \textbf{Middle:} the polarization fraction. \textbf{Bottom:} the distribution of the inclination angle $\gamma$. Contours outline the intensity structures of $^{12}$CO starting from 5 K km/s.}
    \label{fig:gamma}
\end{figure}

\section{Results}
\label{sec:results}
\subsection{Gravitational contraction and collapse}
The PDF of H$_2$ column density is presented in Fig.~\ref{fig:NH2}. The PDF appears as log-normal at a low-density range, while it becomes power-law at a high-density range. In particular, the power-law part exhibits two distinct slopes. A shallow slope $\approx-0.99$ is observed in the range of $0.35\le s\le1.51$, while the slope becomes steep for $s\ge1.51$. The pow-law distribution indicates the presence of a self-gravitating medium while its slope characterizes the star formation activity. Slope $\approx -1$ suggests a rapid collapse process associated with a radial density distribution $\rho(r) \propto r^{-3}$, where $\rho$ is volume mass density and $r$ is the radius to the center of collapse \citep{2013ApJ...763...51F}. Slope $\approx -2.5$ suggests $\rho(r) \propto r^{-2}$ in approximation. The two different slopes also results in two transition density $S_t\approx0.35$ and $S_t\approx1.51$ corresponding to column density $N_{\rm H_2}\approx5.5\times10^{21}$ cm$^{-2}$ and $N_{\rm H_2}\approx1.7\times10^{22}$ cm$^{-2}$, respectively. We outline the density structures within these two density ranges in Fig.~\ref{fig:NH2}. The high-density structures $N_{\rm H_2}\ge1.7\times10^{22}$ cm$^{-2}$ are more filamentary than the relatively low-density structures $5.5\times10^{21}$ cm$^{-2}$ $\le N_{\rm H_2}\le1.7\times10^{22}$ cm$^{-2}$. While both of the structures are identified as gravity-dominated, it is more likely that the low-density structures are under global gravitational contraction at scale $\approx 1$ pc and high-density structures are gravitationally collapsing or fragmenting at scale $\approx0.2$ pc. The corresponding infalling motion is evident in the increase of NH$_3$'s LOS velocity close to Oph B, C, E, and F \citep{2021A&A...648A.114C}. 


\subsection{POS magnetic field orientation}
\label{subsec:pos-b-field}
Figs.~\ref{fig:plank} and \ref{fig:hawc} show the POS magnetic field morphology traced by Planck and HAWC+ dust polarization, respectively. Planck focuses on a large field-of-view $\approx1.6^\circ\times1.6^\circ$ and large-scale magnetic field $\approx0.2$~pc, while HAWC+ zooms in the dense core Oph A and provides information of small-scale magnetic field $\approx0.003$~pc. We see that the large-scale magnetic field is along L1688's north low-intensity tail. The magnetic field, however, is bent in the surroundings of the high-density clump appearing as an hourglass shape. This bending of the magnetic field happens exactly at the gravity-dominated region identified by the PDF (see Fig.~\ref{fig:NH2}). The hourglass magnetic field morphology is more apparent in the surroundings of the dense core Oph A (see HAWC+ \SI{154}{\micro\meter} in Fig.~\ref{fig:hawc} and \citealt{2019ApJ...882..113S} for a streamline visualization). The magnetic field is dragged into the densest region. 

Apart from dust polarization, we also use VGT to trace the magnetic field morphology. VGT-H I, VGT-$^{12}$CO, and VGT-$^{13}$CO provide a view of large-scale magnetic field (H I: ${\rm FWHM}\sim20'$; $^{12}$CO and $^{13}$CO: ${\rm FWHM}\sim7'$) over the entire L1688 clump, while C$^{18}$O gives the information of small-scale magnetic field (${\rm FWHM}\sim0.8'$) in a zoom-in region. The magnetic fields inferred from VGT-$^{12}$CO and VGT-$^{13}$CO measurements are globally similar. VGT agrees well with the Planck polarization at L1688's north low-intensity tail. Misalignment appears in the central dense clump coincident with the self-gravitating region identified by the PDF (see Fig.~\ref{fig:NH2}). The misalignment, therefore, comes from the effect of self-gravity and it is more significant in the dense-gas tracer $^{13}$CO. The misalignment, however, is usually less than 90$^\circ$ suggesting the gravitational collapse and gravitational contraction is not strong enough at scale $\approx0.2$~pc and volume density $\approx10^3$ cm$^{-3}$ so that the velocity gradient of turbulence is not overwhelmingly dominated by gravitational acceleration. 

Importantly, although VGT-H I measurement is associated with the foreground and background rather than the molecular cloud, the VGT-H I magnetic field in the central dense clump agrees with Planck, VGT-$^{12}$CO, and VGT-$^{13}$CO measurements. This agreement means the variation of the magnetic field is insignificant in the foreground/background and the molecular cloud. The magnetic field in this case is expected to be strong so that the molecular gas' motion is channeled by the magnetic field. The cloud is contracting and accreting gas along the magnetic field direction resulting in a disk-like or filamentary high-intensity clump Oph A, C, E, and F (see Fig.~\ref{fig:NH2}). 

VGT-C$^{18}$O zooms in the small dense clump Oph A. The gradient's orientation in the low-intensity region is almost along the west-east direction, while it changes by $90^\circ$ in the high-intensity region being along the north-south direction. Comparing with HAWC+ polarization, the north-south orientation shows AM = -1, i.e., VGT is perpendicular to the polarization (see Fig.~\ref{fig:AM}). This change of gradient's orientation and perpendicular relative angle both suggest the high-intensity part of Oph A is undergoing rapid gravitational collapse at volume density $\approx10^4$ cm$^{-3}$. Earlier Zeeman splitting measurements from OH or CN observations discovered that the molecular clouds at volume density $\approx10^5-10^8$ cm$^{-3}$ are mainly self-gravitating \citep{2012ARA&A..50...29C}. Our finding shows the volume density threshold for the collapse could be lower. However, more samples are required to achieve a statistically general conclusion.

\begin{figure}
	\includegraphics[width=1.0\linewidth]{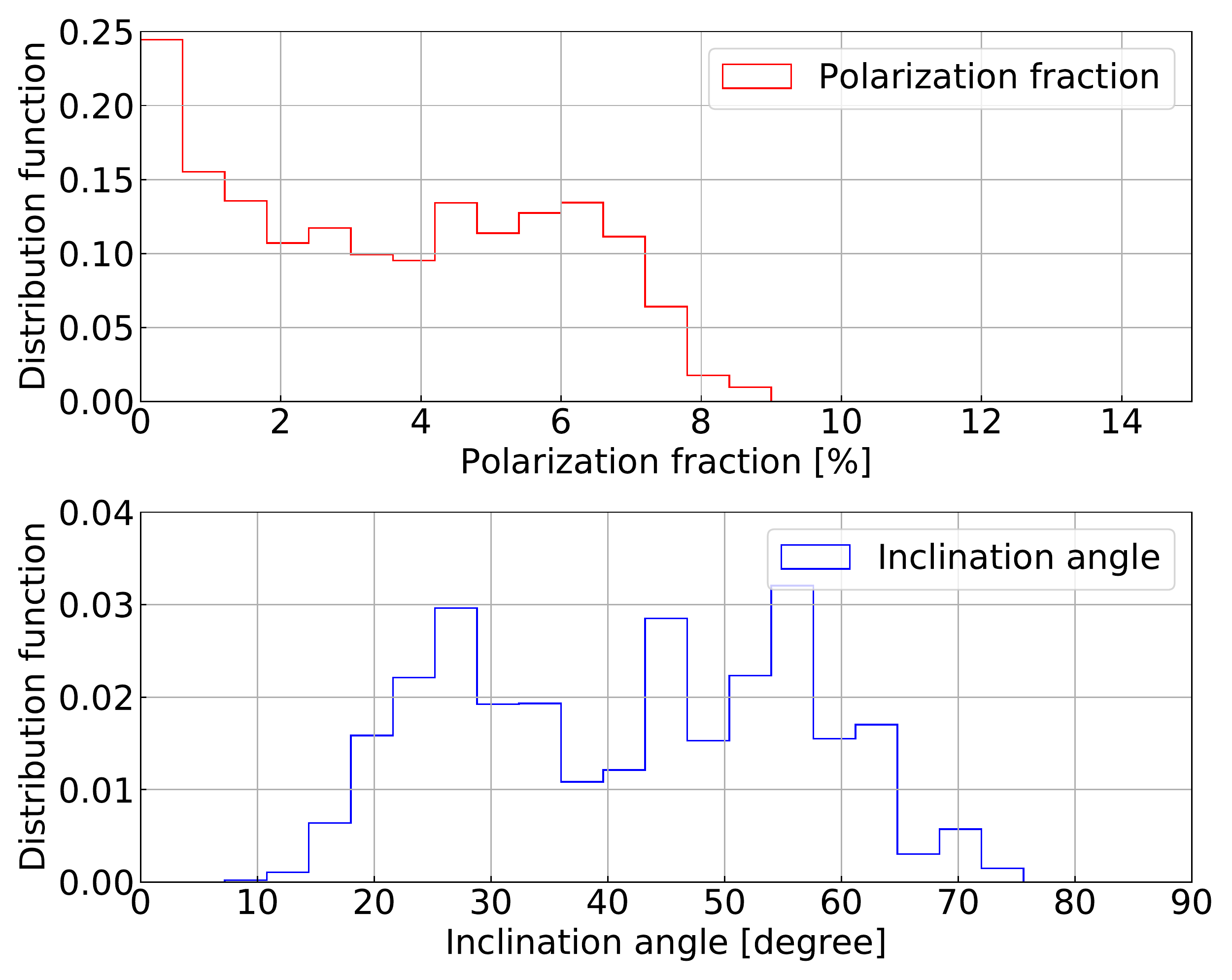}
    \caption{Histogram of Planck's polarization fraction (top) and inferred inclination angle (bottom).}
    \label{fig:gamma_hist}
\end{figure}

\begin{figure}
	\includegraphics[width=1.05\linewidth]{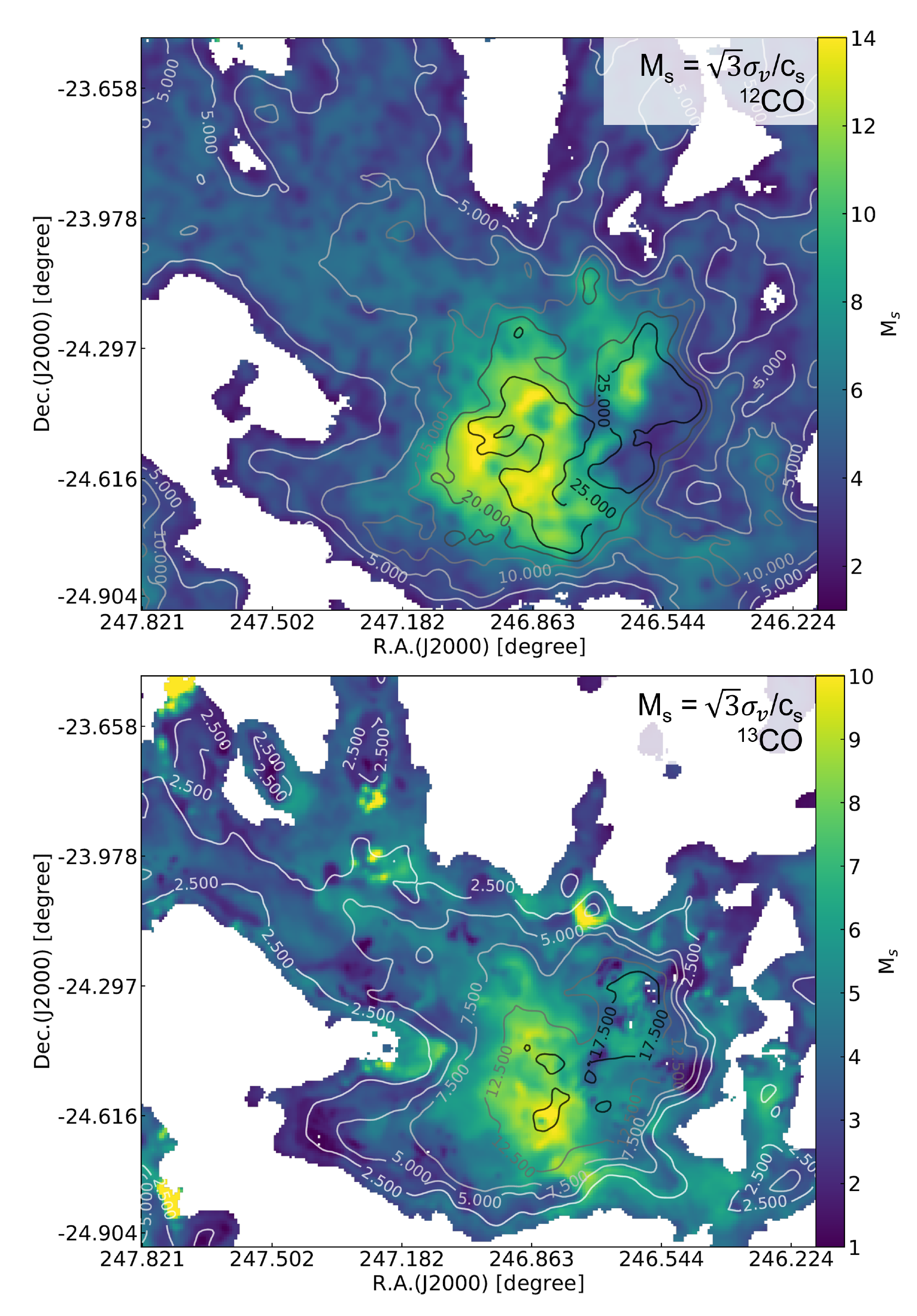}
    \caption{The distribution of sonic Mach number ${\rm M_s}=\sigma_v/c_s$. $\sigma_v$ is calculated from the linewidth of $^{12}$CO (top) and $^{13}$CO (bottom). Contours outline the intensity structures of $^{12}$CO (top) starting from 5 K km/s and $^{13}$CO (bottom) starting from 2.5 K km/s.}
    \label{fig:ms}
\end{figure}

\begin{figure}
	\includegraphics[width=1.05\linewidth]{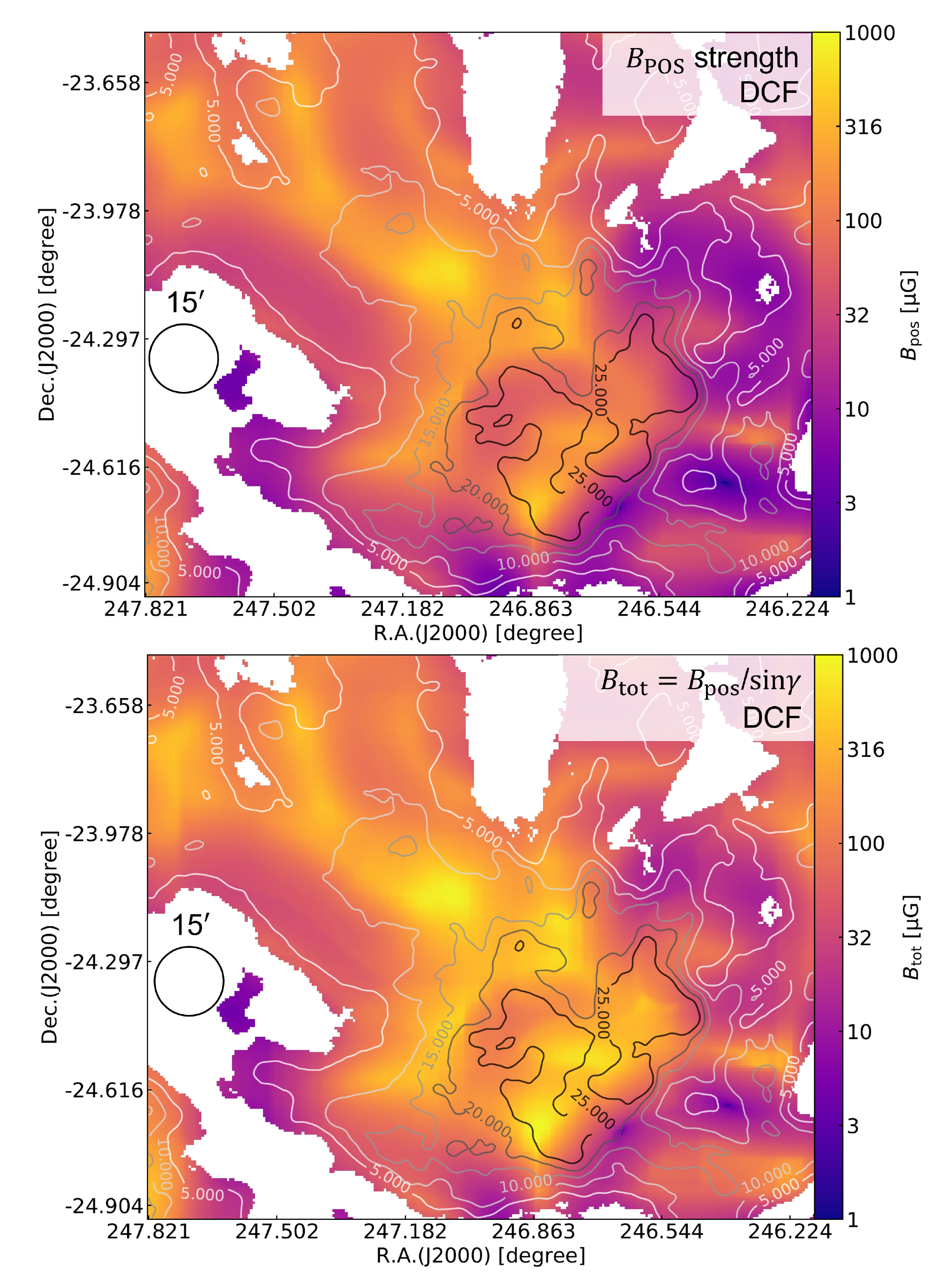}
    \caption{\textbf{Top:} the distribution of the POS magnetic field strength calculated from the DCF method. \textbf{Bottom:} the distribution of the total magnetic field strength $B_{\rm tot}$. Contours outline the intensity structures of $^{12}$CO starting from 5 K km/s.}
    \label{fig:dcf}
\end{figure}

\begin{figure}
	\includegraphics[width=1.05\linewidth]{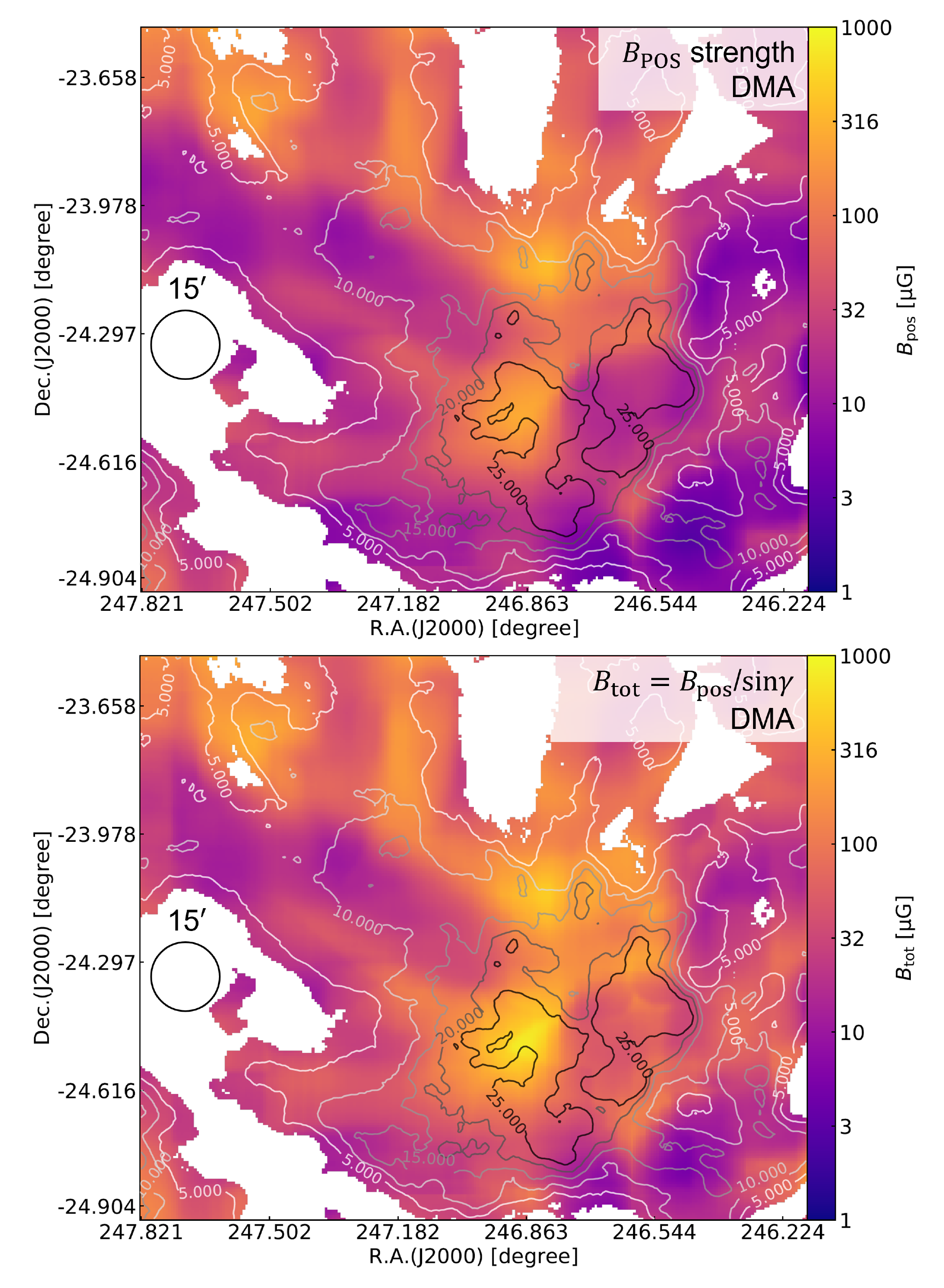}
    \caption{\textbf{Top:} the distribution of the POS magnetic field strength calculated from the DMA method selecting $l=10$~pixels. \textbf{Bottom:} the distribution of the total magnetic field strength $B_{\rm tot}$. Contours outline the intensity structures of $^{12}$CO starting from 5 K km/s.} 
    \label{fig:dma}
\end{figure}

\begin{figure}
	\includegraphics[width=1.0\linewidth]{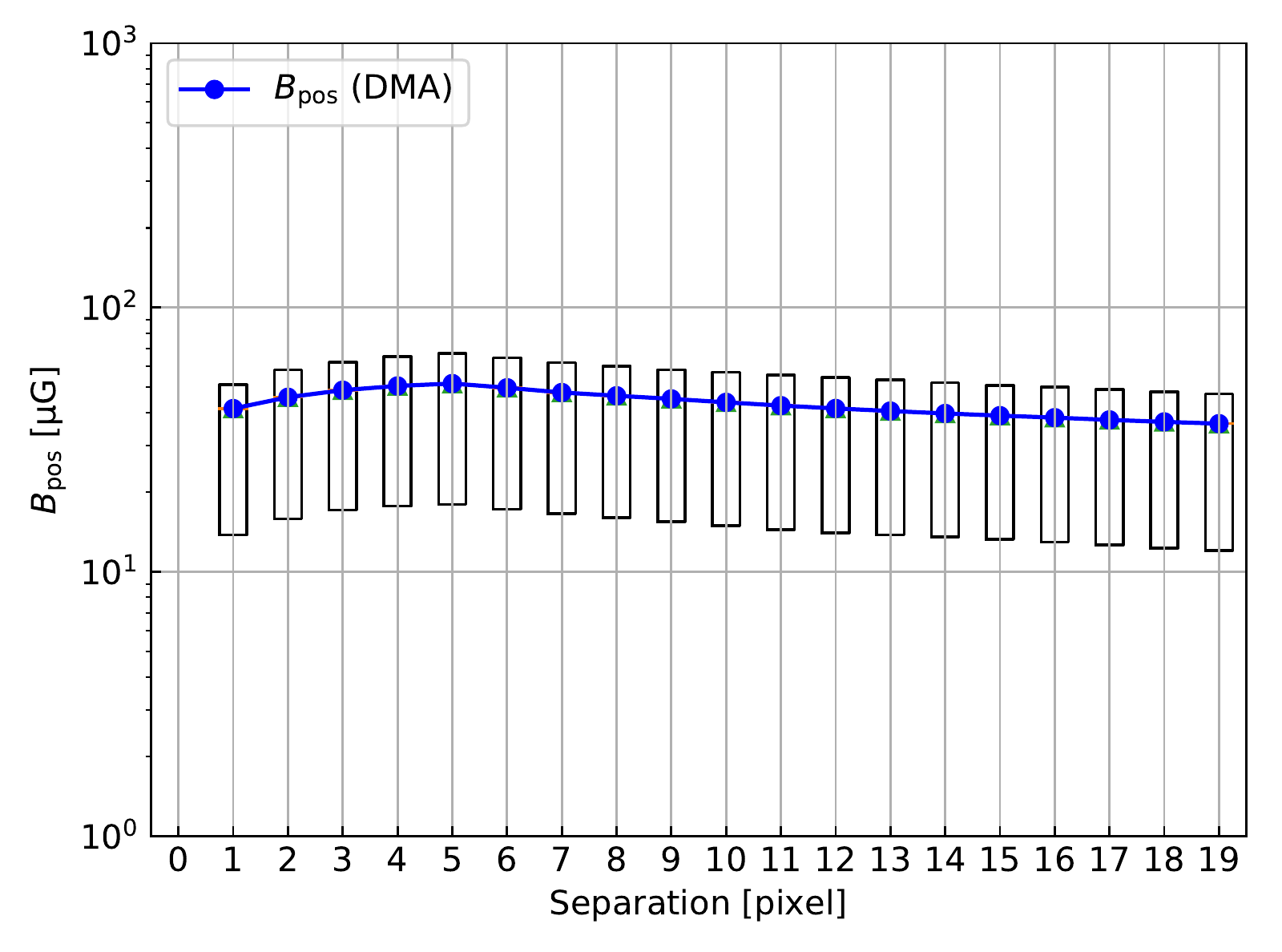}
    \caption{The global mean value of the POS magnetic field strength calculated from the DMA method as a function of the separation $l$. Box gives ranges of the first (lower) and third quartiles (upper) and colored line represents the mean value. }
    \label{fig:dma_L}
\end{figure}

\subsection{Three-dimensional magnetic field orientation}
The calculation of $\gamma$ relies on the dispersion of polarization angle and polarization fraction, which are displayed in Fig.~\ref{fig:gamma}. The northwest part of L1688 shows small-angle dispersion and a relatively high polarization fraction. However, the dispersion dramatically increases in the dense southwest part and the polarization fraction decreases to $\approx2\%$. The low polarization fraction suggests that the large dispersion results from a small inclination angle so that the POS component is weak and more turbulent. 

The distribution of $\gamma$ is presented in Fig.~\ref{fig:gamma} and the histograms of polarization fraction and inclination angle are given in Fig.~\ref{fig:gamma_hist}. The inclination angle of the northeast tail is around $45^\circ-70^\circ$, but it changes to $10^\circ-40^\circ$ in the southwest clump. The Planck polarization reported a maximum polarization fraction $\approx22\%$ across the full sky \citep{2020A&A...641A..12P}, which should correspond to the case of inclination angle $\approx90^\circ$. 

For our data we have the maximum fraction $p_{\rm max}\approx10\%$ due to the depolarization effect caused by the changes in the magnetic field's direction within the beam and along the LOS \footnote{ Note that the condition of $\sigma_\phi^2\ll1$ used in PFA means the magnetic fluctuations are minimum in that position, so the depolarization is dominantly caused by the mean inclination angle.}. To find out the uncertainty in obtained inclination angle due to the beam-average effect, we repeat the analysis for the HAWC+ zoom-in region presented in Fig.~\ref{fig:plank}'s middle panel. Polarization in this region is measured at wavelength \SI{89}{\micro m} and \SI{154}{\micro m} (similar to Planck 353 GHz being submillimetre-wavelength), with beam resolution $\approx7.8''$ and $\approx13.6''$, respectively, so that the beam average effect is reduced. For \SI{154}{\micro m}, we get $p_{\rm max}\approx20\%$ and $p_{\rm off}\approx14\%$, which is averaged over the polarization fraction in the five positions associated with the smallest five $\sigma_\phi$. Consequently, we get $\gamma\approx58^\circ$, which is a bit larger than that ($\approx40^\circ$) derived from the Planck polarization. The difference, although not significant, does not necessarily mean the beam average effect vanishes, especially Planck's $p_{\rm max}\approx10\%$ is only half of the one obtained from HAWC+. We expected the beam average effect appears in both Planck's $p_{\rm off}$ and $p_{\rm max}$ so that its contribution in Eq.~\ref{eq.gamma}'s the numerator and denominator partially cancels off. Moreover, HAWC+ is measuring the magnetic field at smaller scale and denser regions, it is also possible the loss of grain alignment at high-density regions and local variation of the magnetic field contribute to the difference and uncertainty. We discuss it more in \S~\ref{subsec:uncer}.

\subsection{Velocity dispersion and mass density}
The velocity dispersion $\sigma_v$ maps measured from $^{12}$CO and $^{13}$CO are presented in Fig.~\ref{fig:ms}. $\sigma_v$ is calculated from the Full Width at Half Maximum (FWHM): $\sigma_v={\rm FWHM}/2.355$. We identify the central peak of the spectrum and the velocities corresponding to half of the central peak. The velocity difference is taken as the FWHM. We note that $^{12}$CO is typically optically thick and subject to self-absorption, which can result in a double-peaked line, which may raise a bit of overestimation in FWHM, or complicated velocity components. For the latter, we use only the most prominent component, as it represents the maximum turbulence. It is worth noting that $^{12}$CO is also sensitive to outflows, which can increase the velocity dispersion. Since we did not exclude the contribution of outflows, caution should be exercised in regions with significant outflows, such as VLA 1623 in Oph A and IRS 45/47 in Oph B \citep{2006AJ....131.2921R,2015MNRAS.447.1996W}.

By assuming the excitation temperature can be approximated by dust temperature $T_{\rm ex}=T_{\rm dust}$, we convert the velocity dispersion to sonic Mach number $M_{\rm s}=\sqrt{3}\sigma_v/c_s$ using the Herschel dust temperature map \citep{2020A&A...638A..74L}. Here $c_s=\sqrt{\frac{T_{\rm ex}k_B}{m_{\rm H}\mu_{\rm H_2}}}$ ($k_B$ is Boltzmann constant, $m_{\rm H}=1.67\times10^{-24}$~g is the mass of a hydrogen atom, and $\mu_{\rm H_2}=2.8$ is mean molecular weight, see \citealt{2008A&A...487..993K}) is the isothermal sound speed. We can see the central clump in the surrounding of Oph B, E, and F is highly supersonic with peak $M_{\rm s}\approx15$ for $^{12}$CO and $\approx10$ for $^{13}$CO. The ambient gas is still supersonic but has smaller $M_{\rm s}$ values. The global mean $M_{\rm s}$ is around 5 for $^{12}$CO and around 4 for $^{13}$CO. The contribution from thermal speed to $\sigma_v$ is therefore insignificant for such a highly supersonic cloud. 

The volume mass density $\rho$ is calculated from $\rm H_2$ column density. As the central dense clump is gravitationally contracting and its POS projection is approximately circular, naturally, we can assume the L1688 is as deep as it is wide. The effective diameter is $L=2\sqrt{A/\pi}$, where $A$ is the area within $^{13}$CO's intensity contour of 5 K km/s (see Fig.~\ref{fig:ms}) corresponding to the gravity-dominated region identified by the PDF and VGT (see Fig.~\ref{fig:NH2}). This contour well represents the clump's POS projection as it covers the majority of the clump and part of the northeast tail. Here we get $L\approx1.84$~pc and calculate the mass density map from $\rho=N_{\rm H_2}\mu_{\rm H_2}m_{\rm H}/L$.


\subsection{Total magnetic field strength}
\subsubsection{3D Magnetic field strength via the DCF}
For estimating the POS magnetic field strength over a region of interest, the DCF uses the velocity dispersion estimated from $^{12}$CO's line broadening, the dispersion of magnetic field angle obtained from polarization's angular statistics, and the mass density calculated from $\rm H_2$ column density. These maps are presented in Figs.~\ref{fig:NH2}, \ref{fig:gamma_hist}, and \ref{fig:ms}. Particularly, the mass density map and $^{12}$CO's velocity dispersion map are smoothed by the Gaussian filter to achieve the same resolution as Planck polarization's dispersion map. The choice of $^{12}$CO is based on the fact that VGT-$^{12}$CO gives the best agreement with Planck polarization suggesting the self-gravity effect is minimum. The obtained POS magnetic field strength map is shown in Fig.~\ref{fig:dcf}. The POS magnetic field in the low-density northeast tail is stronger than the one in the central dense clump. However, this is caused by the projection effect. 

We recover the total magnetic field strength by using the inclination angle. Without the projection effect, the central dense clump consequently exhibits the strongest total magnetic field due to compression and gravitational collapse. The maximum strength achieves $\approx$~\SI{1100}{\micro G}, and the global mean value is $\approx$~\SI{135}{\micro G}.

\subsubsection{3D Magnetic field strength via DMA}
Different from the DCF method, DMA uses the second-order structure function to calculate the velocity dispersion from the velocity centroid map and the magnetic field angle dispersion from the polarization map. The DMA-estimated POS magnetic field strength distribution is presented in Fig.~\ref{fig:dma}. Here we select $l=10$~pixels for calculating the structure-function (see Eq.~\ref{eq.dma}). As we show in Fig.~\ref{fig:dma_L}, the DMA estimation is insensitive to $l$. While variation exists, the global mean values of the POS component ($\approx$~\SI{40}{\micro G} - \SI{50}{\micro G}), as well as the distributions, for different $l$ values are statistically similar. The slightly decreasing trend may be contributed by non-turbulent fields when considering large-scale fluctuations. In combination with the inclination angle distribution, we see the mean value of 3D magnetic field strength at $l=10$~pixel is $\approx$ \SI{75}{\micro G}, which is smaller than the DCF estimation. 

Moreover, we also observe other apparent differences compared with the DCF estimation (see Fig.~\ref{fig:dcf}). In the north tail and central clump, the DMA-estimated magnetic field strength in high-density (relative to its surroundings, see Fig.~\ref{fig:NH2}) regions, is weaker. In general, we expect the effects of the self-gravity to distort the magnetic field direction and decrease the value of the magnetic field. In the high-density Oph A region, the magnetic field bending caused by self-gravity increases the polarization angle's dispersion, but the DCF method used in this work does not account for such an effect \footnote{ Note in other modified DCF methods, the effect of field line distortions whether due to self-gravity or otherwise are considered \citep{2009ApJ...696..567H,2009ApJ...706.1504H}. This effect can be handled by DMA that uses structure function to calculate the angle dispersion \citep{2022arXiv220409731L}. }. As a result, we expect that Oph A gives excessive angle dispersion, which results in underestimating the actual magnetic field strength (see \S~\ref{subsec:uncer} for more discussion about uncertainty). However, DCF may intrinsically have overestimation (see \S~\ref{sec:dis}) so that the self-gravity effect is only apparent in DMA.

\begin{figure}
	\includegraphics[width=1.0\linewidth]{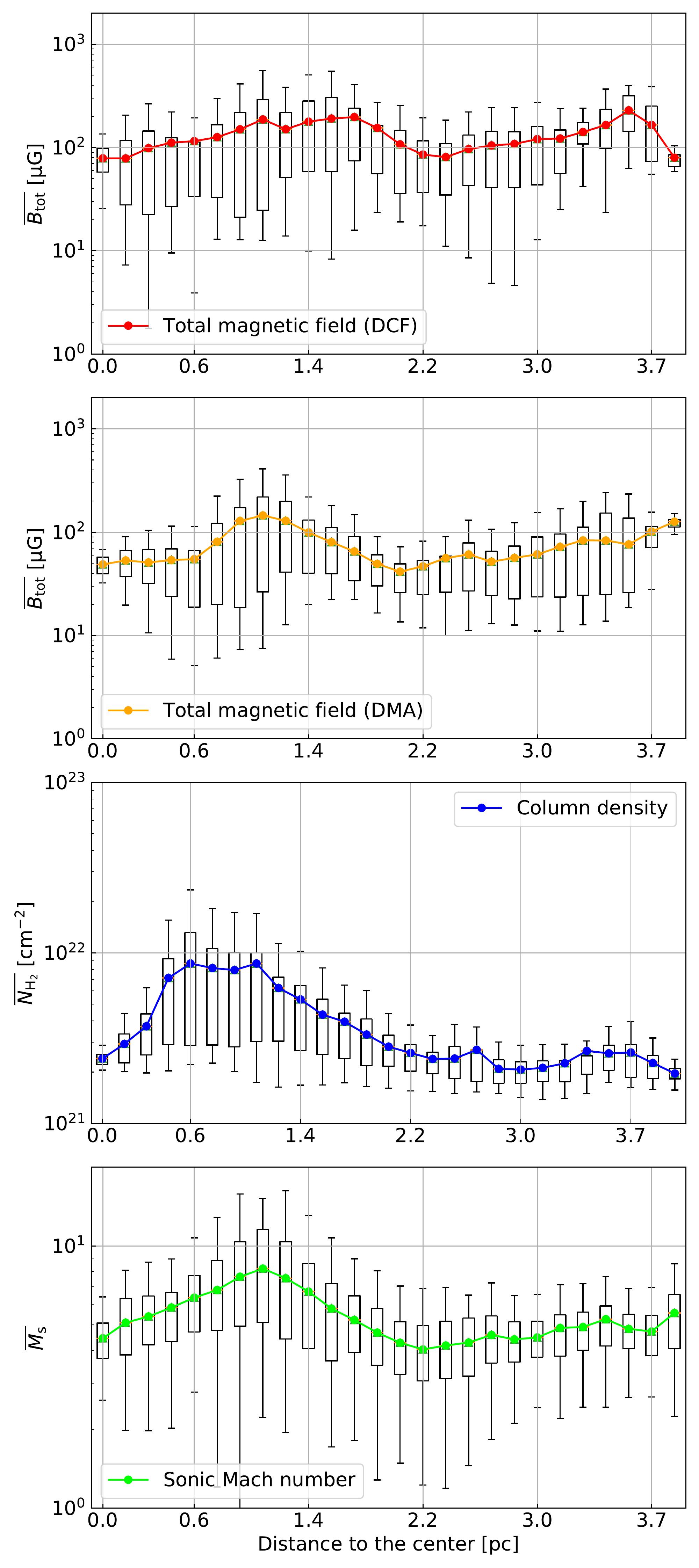}
    \caption{The annulus-averaged total magnetic field estimated from DCF (the first panel), total magnetic field estimated from DMA (the second panel)  column density (the third panel), and sonic Mach number calculated from $^{12}$CO (the fourth panel) as a function of the distance to the cloud's center. The width of each annulus is $\approx0.15$~pc. The upper and lower black lines represent the maximum and minimum values, respectively. Box gives ranges of the first (lower) and third quartiles (upper) and the colored line represents the mean value. }
    \label{fig:Bring}
\end{figure}

\begin{figure}
	\includegraphics[width=1.0\linewidth]{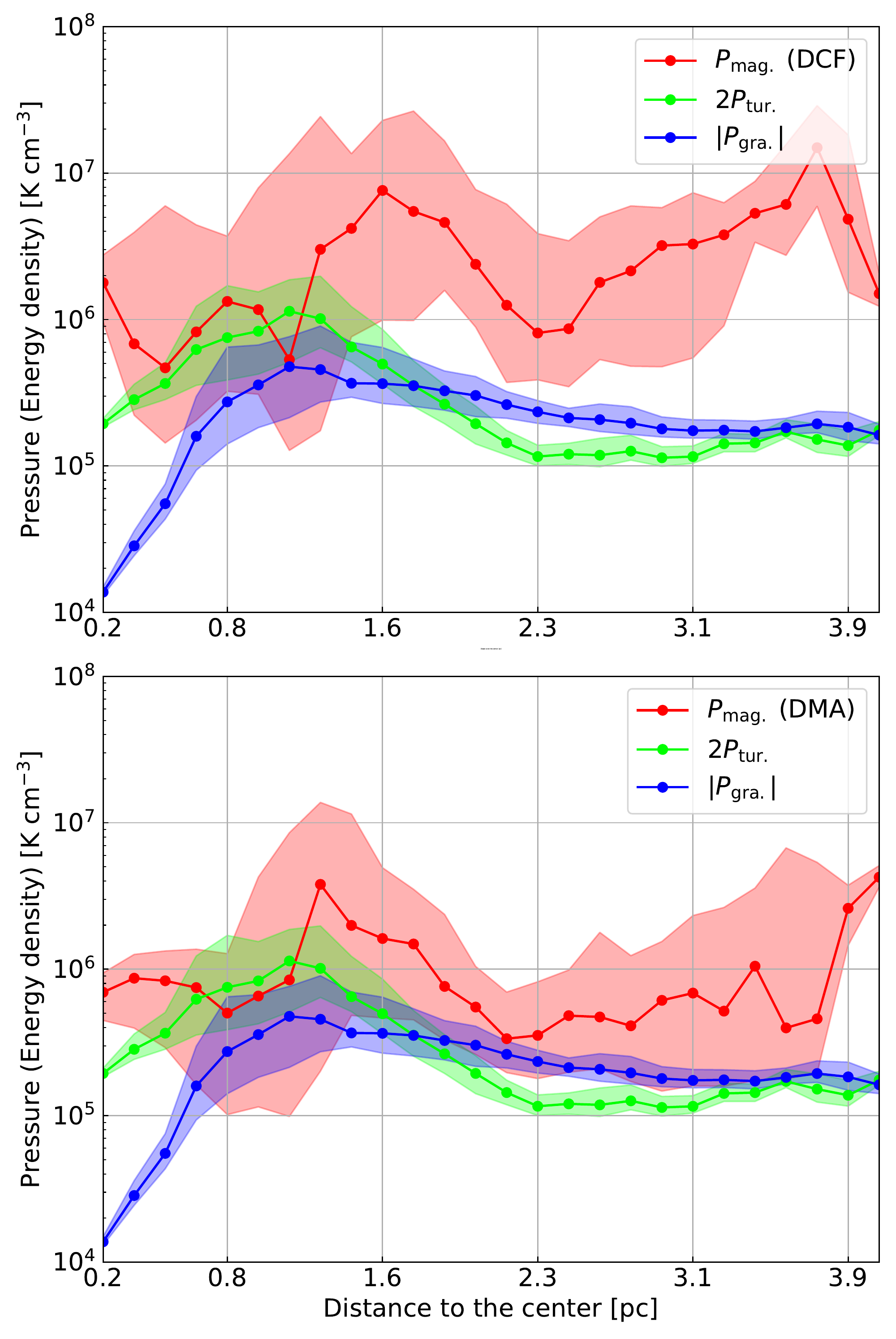}
    \caption{The annulus-averaged pressure (i.e., energy density) of magnetic field (red; top: from DCF; bottom: from DMA), turbulence (green), and gravitational potential (blue) as a function of the distance to the cloud's center. The width of each annulus is $\approx0.15$~pc. The Shadow area ranges from the first (lower) to third quartiles (upper) and the colored line represents the mean value.}
    \label{fig:pressure}
\end{figure}

\begin{figure}
	\includegraphics[width=1.0\linewidth]{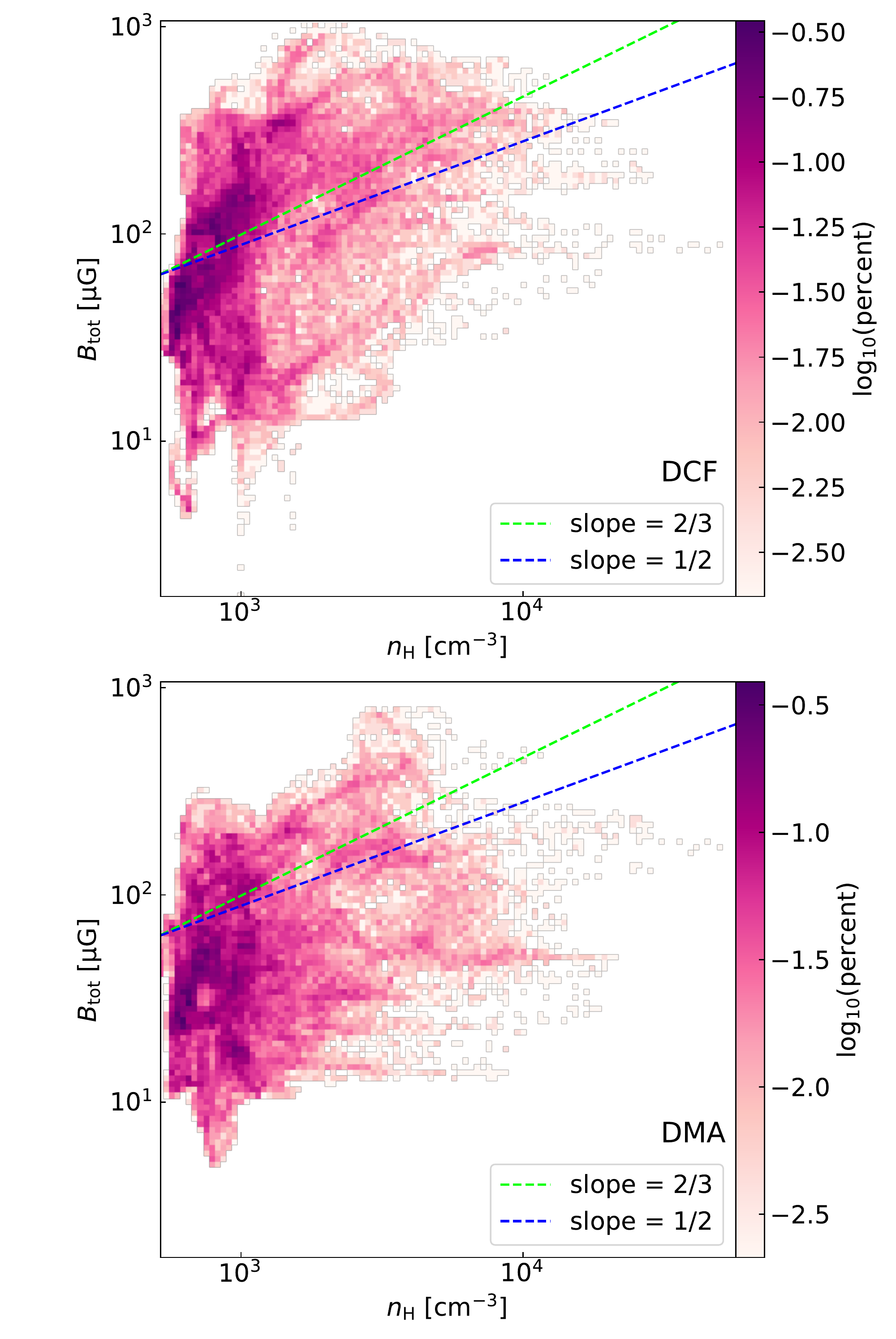}
    \caption{2D histogram of logarithmic volume density $n_{\rm H}$ and logarithmic total magnetic field strength estimated from DCF method (top) or DMA technique (bottom). The slopes are overplotted for comparison.}
    \label{fig:Bvsnh2}
\end{figure}

\begin{figure}
	\includegraphics[width=0.99\linewidth]{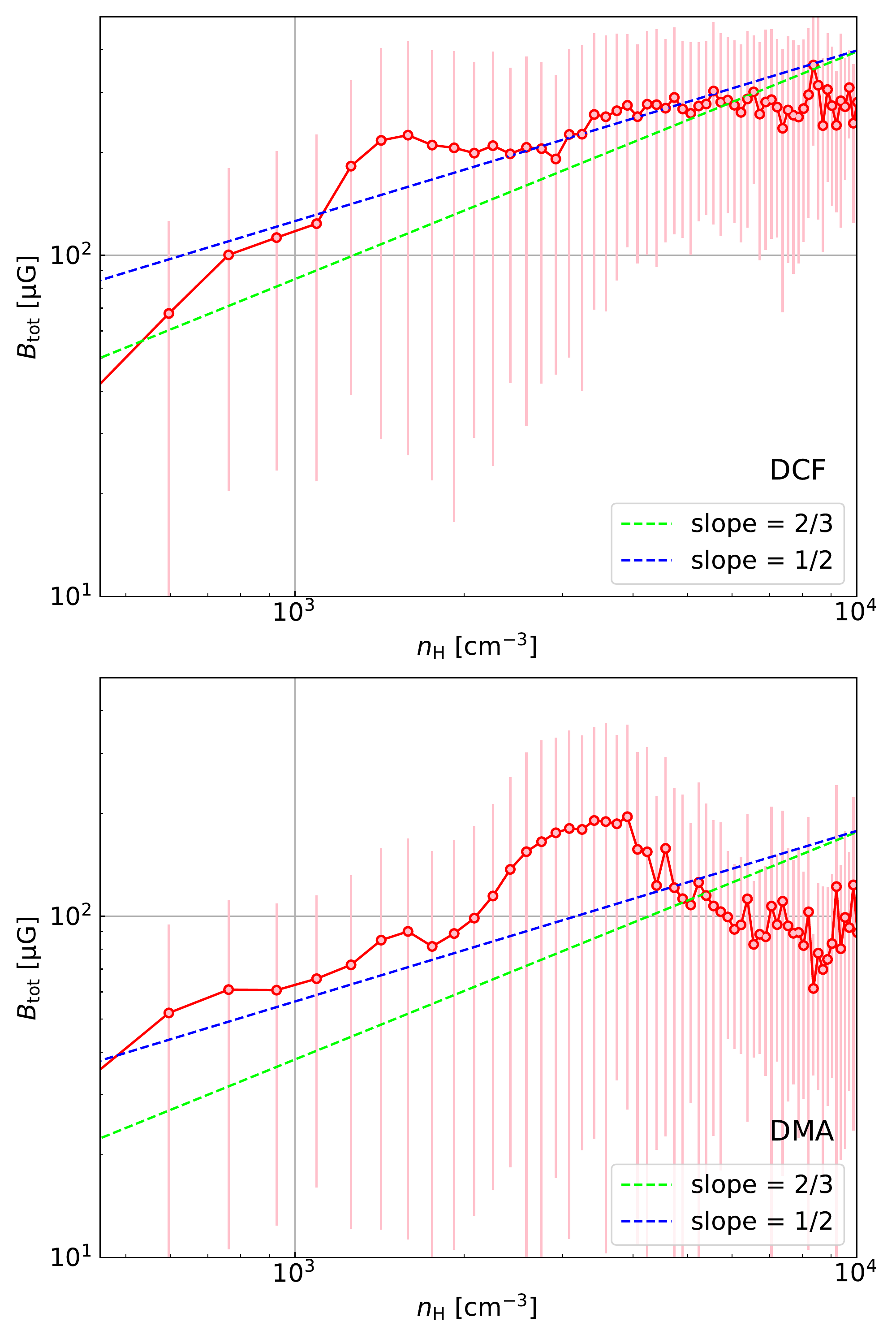}
    \caption{The correlation of volume density $n_{\rm H}$ and total magnetic field strength estimated from the DCF method (top) or DMA technique (bottom). The magnetic field strength is averaged over uniformly spaced (volume density) bins. The error bar is given by the standard deviation. The slopes are overplotted for comparison.}
    \label{fig:Bvsnh2_bin}
\end{figure}

\subsection{Energy budget}
To quantify how the magnetic field, turbulence, and self-gravity interact in  L1688, we first assume the cross point (R.A.$\approx246.65^\circ$, and Dec.$\approx-24.52^\circ$) of Oph A and B's elongation as the collapsing center (see Fig.~\ref{fig:NH2}). Then we take the annulus average for the three quantities as a function of the distance to the center. The width of each annulus is $\approx0.15$~pc. As shown in Fig.~\ref{fig:Bring}, the magnetic field, $M_{\rm s}$, and column density have a significantly higher value in the vicinity ($\approx0.6-1.5$~pc) of the collapse center, while they dramatically decrease in the envelope of the cloud ($\approx2.0-3.0$ pc). The increment around the center can be easily understood because the gravitational contraction/collapse accretes gas and compresses the magnetic field. The infall velocity and gravity-driven turbulence also contribute to the velocity dispersion resulting in a high sonic Mach number.

We similarly take calculate the annulus-averaged energy density (i.e., pressure) of the magnetic field, turbulence, and gravitational potential as a function of the distance to the cloud's center. The magnetic and isotropic turbulent pressure could be computed as:
\begin{equation}
\begin{aligned}
    P_{\rm mag.}&=\frac{B_{\rm tot}^2}{8\pi},\\
    P_{\rm tur.}&=\frac{3}{2}\rho\sigma_v^2.\\
\end{aligned}
\end{equation}
As for the pressure of gravitational potential, we can consider that a mass element $\rho(r)dV$ (here $V$ represents volume) at radius $r$ is attracted by a central massive sphere. The pressure is then:
\begin{equation}
    P_{\rm gra.}=-\frac{GM(r)}{r}\rho(r),\\
\end{equation}
where $G$ is the gravitational constant and $M(r)$ is the sphere's mass within radius $r$. $\rho(r)$ is taken from the annulus-averaged mass density.

As shown in Fig.~\ref{fig:Bvsnh2}, the magnetic field gives the strongest supportive pressure $\approx10^7$ K cm$^{-3}$ for DCF and $\approx5\times10^6$ K cm$^{-3}$ for DMA. Although the turbulence and gravitational potential pressure increases in the center's vicinity ($\approx0.5-1.5$~pc), their magnitudes are at a comparable level to the magnetic field's pressure. In particular, instead of viral equilibrium, we observed $|P_{\rm gra.}|>2P_{\rm tur.}$ due to the fact that the magnetic field acts against gravity so that the velocity dispersion is expected to be lower in the magnetized cloud. However, $|P_{\rm gra.}|$ is slightly larger than $2P_{\rm tur.}$ when the distance is larger than $\approx1.5$~pc and $2P_{\rm tur.}$ exceeds $|P_{\rm gra.}|$ in the center's vicinity ($\le1.5$~pc) . The higher $2P_{\rm tur.}$ pressure could be contributed by the infall velocity from gravitational collapse and gravity-driven turbulence suggesting the ambient material is falling into the center. The cloud's averaged Jeans length is $L_J=\sqrt{\frac{\pi c_s^2}{G\overline{\rho}}}\approx0.70$~pc with a mean volume mass density $\overline{\rho}\approx3.52\times10^{-21}$ g~cm$^{-3}$ (i.e., volume number density around 750 cm$^{-3}$). When the distance is less than the Jeans length, thermal support is stronger than self-gravity. Consequently, the global contraction slows down within $\approx0.70$~pc, and local collapse and fragmentation appear.
\subsection{Correlation of total magnetic field strength and volume density}
The correlation between the total magnetic field strength and the volume mass density is an important aspect of the theory of star formation. Under the flux-freezing ideal MHD condition, the scaling $B_{\rm tot}\propto \rho^{2/3}\propto n_{\rm H}^{2/3}$ corresponds to the isotropic gravitational contraction of a sphere \citep{1965QJRAS...6..265M,2001ApJ...562..852H,2011MNRAS.414.2511V}, where $n_{\rm H}=2n_{\rm H_2}\approx2N_{\rm H_2}/L$ is the volume density of hydrogen atom. However, the scaling can be $B_{\rm tot}\propto n_{\rm H}^{1/2}$ if the flux-freezing condition breaks due to reconnection diffusion \citep{2005AIPC..784...42L,2014SSRv..181....1L}. In addition, the initial contraction can preferably take place along the strong magnetic field so that the gas settles into a flattened cylindrical structure perpendicular to the mean field \citep{1976ApJ...207..141M,1980ApJ...239..166S,1999ApJ...520..706C}.

In Fig.~\ref{fig:Bvsnh2}, we show the logarithmic $B_{\rm tot}$-$n_{\rm H}$ scatter 2D histogram. At a low-density regime, the scatter dots cover an extended area without apparent scaling relation. At high densities, the scatter is reduced, but characteristic scaling is still not clearly distinct. The scatter dots lie between scaling with slopes 1/2 and 2/3. We further bin the measured total magnetic field strength in uniformly spaced $n_{\rm H}$ bins with an interval of $\approx150$ cm$^{-3}$, as shown in Fig.~\ref{fig:Bvsnh2_bin}. For both cases of DCF and DMA measurements, when volume density is smaller than $6.0\times10^3$ cm$^{-3}$, the characteristic scaling is more close to $1/2$. This indicates that the collapse is affected magnetic field. The measured turbulence can induce reconnection diffusion or the collapse can preferably take place along the strong magnetic field. However, one should note that it is possible the correlation of $B_{\rm tot}\propto n_{\rm H}^{1/2}$ results from the intrinsic assumption of $B_{\rm pos}\propto\sqrt{4\pi\rho}$ in the DCF and DMA method. 

\section{Discussion}
\label{sec:dis}
\subsection{3D magnetic field in L1688}
Probing the magnetic field in the molecular cloud is always challenging. In this work, we rely on dust polarization and VGT to access the POS magnetic field in the L1688 star-forming region. L1688's low-intensity northeast tail follows the magnetic field direction. The magnetic field, however, is slightly bent in the vicinity of the central dense clump being an hourglass shape (see Fig.~\ref{fig:plank}, HAWC+ \SI{154}{\micro\meter} in Fig.~\ref{fig:hawc}, and \citealt{2019ApJ...882..113S} for a better streamline visualization of HAWC+ \SI{154}.). This hourglass feature is a very specific prediction of gravitational collapse in a region with a strong magnetic field \citep{1993ApJ...417..243G,1993ApJ...415..680F,2011A&A...535A..44F}.

The magnetic fields measured by VGT-$^{12}$CO,  VGT-$^{13}$CO, and VGT-H I generally agree with the one inferred from dust polarization. The agreement reveals the fact that the magnetic field is coherent up to volume density $n_{\rm H_2}\approx10^3$ cm$^{-3}$ at least. In particular, VGT-H I traces the magnetic field associated with the foreground/background rather than the molecular cloud L1688. However, VGT-H I still agrees with VGT-$^{12}$CO, VGT-$^{13}$CO, and Planck polarization suggesting that L1688's motion and evolution are regulated by a strong magnetic field. 

An earlier study of \cite{2021MNRAS.503.5006S} reported an inclination angle $\approx57^\circ$ for the entire Ophiuchus molecular cloud assuming the magnetic field's fluctuation is negligible. In this paper, we employ a new technique employing both polarization fraction and polarization angle dispersion to trace the magnetic field inclination angle with respect to the LOS. We find the magnetic field in L1688's low-intensity northeast tail has an inclination angle $\approx55^\circ$, while the central clump's magnetic field is inclined by $\approx30^\circ$ on average. Our estimation accounts for magnetic field fluctuations and explains the low polarization fraction in L1688. 

\subsection{Dynamics of L1688}
Based on the spectral emissions, $\rm H_2$ column density, and polarization data, we evaluate L1688's dynamical properties. The PDF of $\rm H_2$ column density reveals two different power-law slopes. A shallow slope $\approx-0.99$ is observed in the density range of $5.5\times10^{21}$ cm$^{-2}$ $\le N_{\rm H_2}\le$ $1.7\times10^{22}$ cm$^{-2}$ and a steep slope $\approx-2.49$ appears for $1.7\times10^{22}\le N_{\rm H_2}$ cm$^{-2}$. These two different slopes suggest the L1688 is undergoing global gravitational contraction at large scale $\approx1.0$~pc and gravitational collapse at small scale $\approx0.2$~pc. The observed misalignment of VGT and polarization and the hourglass shape magnetic field morphology confirm this point. In particular, the relative angle of VGT-$^{12}$CO/$^{13}$CO and polarization are less than $90^\circ$ suggesting that turbulence is still more dominated than self-gravity up to the volume density $n_{\rm H_2}\approx10^3$ cm$^{-3}$. However, the $90^\circ$ relative orientation is observed in VGT-C$^{18}$O and HAWC+ polarization in a small dense core ($\approx0.05$~pc) Oph A. Such $90^\circ$ difference reveals a strong gravitational collapse in Oph A at density $n_{\rm H_2}\approx10^4$ cm$^{-3}$.

Moreover, L1688 is highly supersonic and strongly magnetized. The global mean $M_{\rm s}$ is around 7 for $^{12}$CO and around 5 for $^{13}$CO. Its kinetic energy is approximately in viral equilibrium with gravitational potential energy, while the kinetic energy exceeds the gravitational potential energy at a scale less than $\approx1.5$~pc due to the contribution from infall velocity and thermal energy. Based on the DCF method and estimated inclination angle, we find L1688's maximum strength achieves $\approx$~\SI{1100}{\micro G} and the global mean value is $\approx$~\SI{135}{\micro G}. The magnetic field energy is larger than kinetic and gravitational potential energy by more than one order of magnitude. Such a strong magnetic field regulates the gravitational contraction so that we observe a power-law relation $B_{\rm tot}\propto n_{\rm H}^{1/2}$ in the volume density range of  $n_{\rm H}\le6.0\times10^3$ cm$^{-3}$. The reconnection diffusion, possibly assisted by the accumulation of matter along magnetic field lines is most likely the cause of the observed dependence. 

\subsection{Uncertainty}
\label{subsec:uncer}
The major uncertainty in our estimation comes from the DCF method and the polarization fraction method. The DCF method assumes that (i) ISM turbulence is an isotropic superposition of linear Alfv\'en waves, (ii) the compressibility and density variations of the media are negligible, and (iii) the variations of the magnetic field direction and the velocity fluctuations arise from the same region in space. These assumptions could raise uncertainty. In particular, gravitational contraction and collapse can magnify both velocity and the magnetic field's dispersion, resulting in overestimating the POS magnetic field strength. More discussion about the DCF method's uncertainty could be found in \cite{2022arXiv220409731L,2021A&A...656A.118S,2022arXiv220509134C,2022ApJ...925...30L}.

\subsubsection{Uncertainty in the total magnetic field strength}
Another uncertainty may come from our assumption of the cloud's depth along the LOS for calculating the mass density. We assume a spherical cloud model for the calculation. $|P_{\rm gra.}|$ and $2P_{\rm tur.}$ may be underestimated with this assumption. If the cloud is highly sheet-like with a LOS length scale approximately one order of magnitude smaller than $L\approx1.84$~pc (from the spherical model), $|P_{\rm gra.}|$ and $2P_{\rm tur.}$ could be comparable or greater than $P_{\rm mag.}$ from DCF (for DMA measurement, $P_{\rm mag.}$ is approximately four times lower). As the observed POS projection of the cloud is close to circular (see Fig.~\ref{fig:NH2}), the cloud can be sheet-like if it is compressed along the LOS direction. However, the magnetic field's inclination angle is $\approx30^\circ$ (see Fig.~\ref{fig:gamma_hist}) and the insignificant variation of the magnetic field in the foreground/background and the cloud implies a strong magnetic field (see \S~\ref{subsec:pos-b-field}). The cloud can be sheet-like if it is compressed along the LOS direction. We do not expect such a compression almost along the strong magnetic field. In any case, the cloud's 3D real structure potential can be better determined by combining extinction maps and star distance provided by the Gaia mission \citep{2021ApJ...919...35Z}.


Earlier Zeeman measurements found that the LOS magnetic field strength in the H I self-absorption region in front of the cloud is around \SI{10}{\micro G} \citep{1994ApJ...424..208G, 1996ApJ...471..302T}. The Zeeman effect from dense tracer OH, however, is poorly sampled and we can expect the LOS magnetic field in the cloud to be stronger than \SI{10}{\micro G}. Nevertheless, \cite{2012ARA&A..50...29C} statistically reported the upper limit of the LOS magnetic field strength is $\approx$~\SI{100}{\micro G} for volume density $n_{\rm H}\approx1.0\times10^4$ cm$^{-3}$ using Zeeman splitting measurement. Assuming a simple equivalent of the POS and LOS magnetic fields, the DCF-measured total magnetic field strength ($\approx$~\SI{250}{\micro G} for $n_{\rm H}\approx1.0\times10^4$ cm$^{-3}$, see Fig.~\ref{fig:Bvsnh2_bin}) might be overestimated by a factor of 2, but the DMA measurement is more close to the Zeeman results. This overestimation may come from the limited sub-block size selected for calculating the angle dispersion. The velocity dispersion estimated from line broadening corresponds to the velocity fluctuation at injection scale $L_{\rm inj}$, but the POS magnetic field angle's dispersion for a small sub-block with size $s$ corresponds to the fluctuation at scale $s$. Compared to the angle dispersion for the entire cloud (i.e., at turbulence injection scale $L_{\rm inj}$), the dispersion for a small sub-block is reduced by a factor of $\sim(s/L_{\rm inj})^{1/3}$, assuming Kolmogorov-type turbulence. An additional decrease comes from the fact that the observed angle dispersion is a LOS-averaged quantity. This average for a small sub-block suppresses magnetic field angle dispersion by an additional factor $(s/L_{\rm inj})^{1/2}$ in a random walk manner. Thus the angle dispersion is totally underestimated by a factor of $(s/L_{\rm inj})^{5/6}$ and the POS magnetic field is, therefore, overestimated. This overestimation results from a combination of the intrinsic typical DCF overestimation of field strength, and of the effects of the approaches to estimating velocity dispersion and magnetic field angle dispersion \citep{2022arXiv220409731L}.

\subsubsection{Uncertainty in the inclination angle}
The polarization fraction analysis assumes that the intrinsic polarization fraction $p_0$ is constant throughout a cloud. This implicitly requires that dust grains’ properties, i.e., emissivity, temperature, etc., are homogeneous within the cloud. Such assumption would result in a systematic uncertainty of $0-20^\circ$ with a median value of $\sim10^\circ$ \citep{2022arXiv220309745H}. On the other hand, this method requires $\sigma_\phi^2\sim M_{\rm A,POS}\ll1$ in a strongly magnetized reference position. This condition may not be satisfied in the central dense clump as the observed $\sigma_\phi$ is large. However, as numerically shown in \citet{2022arXiv220309745H}, the condition of $M_{\rm A,POS}^2\ll1$ can be relaxed to $M_{\rm A,POS}^2\approx1$ when the inclination angle is less than $30^\circ$. Such a small inclination angle ubiquitously dominates the depolarization effect so that it reduces the contribution from $M_{\rm A,POS}^2$. Therefore, we expect the uncertainty from the condition of $M_{\rm A,POS}^2\ll1$ is not significant in L1688.

On the other hand, polarization fraction is typically strongly anti-correlated with Stokes $I$, which is an effect usually due to the loss of grain alignment at high extinction $A_V$ \citep{2008ApJ...674..304W,2014A&A...569L...1A,2015AJ....149...31J} or field variation along the LOS \citep{2015A&A...576A.105P,2019MNRAS.482.2697S}, a phenomenon known as the "polarization hole". In Fig.~\ref{fig:Ivsgamma}, we present the 2D histogram of the inclination angle $\gamma$ and $I$ and find $\gamma$ tends to be small with a high $I$. However, we note that a sharp drop of polarization fraction happens at hydrogen column density $N_{\rm H}\sim10^{22}$ cm$^{-2}$ (see \citealt{2015A&A...576A.104P}). Our analysis mostly covers low-density regions ($N_{\rm H}<10^{22}$ cm$^{-2}$, see Fig.~\ref{fig:NH2}). In Fig.~\ref{fig:gamma}, we see the large dispersion of polarization angle and small inclination angle also appear in low-density regions (see also the 2D histograms in Fig.~\ref{fig:pdisNH2}). It suggests that the magnetic field’s variation is indeed large. In addition, \cite{2019ApJ...880...27P} studied high-density Oph A, Oph B, and Oph C and found that dust grains may remain aligned with the magnetic field. Considering the fact that the number of aligned large grains decreases in high-density regions \citep{2021ApJ...908..218H}. The grain alignment, in low-density regions, should be better than those observed in Oph A, B, and C. For these two reasons, we, therefore, expect the depolarization due to field variation dominates L1688, especially in low-density regions.

However, the extent to which loss of grain alignment contributes to the polarization hole at parsec scales remains contentious. For instance, \cite{2019MNRAS.482.2697S} performed a numerical study showing that dust grains remain well aligned even at high volume densities ($n > 10^3$ cm$^{-3}$ and $A_V>1$). A more recent study of \cite{2021ApJ...908..218H} estimated the maximum $A_V$ that grains still align with magnetic fields. They found for gas density of $n \approx 10^4$, $10^5$, $10^6$ cm$^{-3}$, the maximum $A_V$ for grain alignment is $\sim20.3, 8.2, 3.3$ for the standard maximum size of interstellar dust $\sim$ \SI{0.25}{\micro\meter}. These values of $A_V$ further increase in the presence of radiation from stars. For our targets, the maximum volume density is around $10^4$ cm$^{-3}$ (see Fig.~\ref{fig:Bvsnh2_bin}) and the maximum $A_V$ is around 20 by adopting a linear relation between the hydrogen column density and $A_V$ ($N_{\rm H} = 2.21\times10^{21}A_V$; \citealt{2009MNRAS.400.2050G}). Nevertheless, for high-density regions, like Oph A, Oph B, and Oph C, the loss of grain alignment can be a potential bias in the PFA analysis. 

\subsection{Comparison with other studies}
The magnetic fields within the Ophiuchus cloud have been extensively studied across different wavelengths. Early optical and near-infrared polarimetry techniques revealed magnetic fields in regions of low extinction \citep{1977AJ.....82..198V, 1979AJ.....84..199W,1988MNRAS.230..321S}. Measurements at high-extinction regions are obtained through far-infrared polarimetry. For example, the JCMT BISTRO POL-2 Survey, as summarized in \cite{2019ApJ...880...27P}, measured the POS magnetic field orientation for Oph A \citep{2018ApJ...859....4K}, Oph B \citep{2018ApJ...861...65S}, and Oph C \citep{2019ApJ...877...43L} at \SI{850}{\micro\meter}, as well as HAWC+ observations at 53, 89, and \SI{154}{\micro\meter} \citep{2019ApJ...882..113S}. In addition, \cite{2018ApJ...859....4K} used the DCF method to derive the POS magnetic field strengths for Oph A cores (0.2 pc), which were found to range from 0.2 to 5 mG. Similarly, \cite{2019ApJ...877...43L} found the POS magnetic field strength in Oph C to be approximately 0.1-0.2 mG. In our analysis, we utilized Planck polarization to derive the magnetic field strength at a larger scale ($\approx6$ pc). Our results generally agree with those of the low-density Oph C, but are smaller than those of the dense Oph A. This outcome is expected as magnetic field strength typically increases in smaller, denser cores, especially when self-gravity dominates, as suggested by \cite{2015MNRAS.450.1094P} and our VGT analysis. In any case, JCMT polarization has a higher resolution of 0.01 pc compared to Planck. The PFA and DMA analysis methods used in this study can be applied to JCMT observations to obtain 3D magnetic field information at small scales.

\section{Summary}
\label{sec:con}
In this work, we study L1688's 3D magnetic field and dynamics. We employ the data sets of spectral emissions of $^{12}$CO, $^{13}$CO, C$^{18}$O, and H I, Planck 353 GHz and HAWC+ \SI{89}{\micro\meter} \& \SI{53}{\micro\meter} polarized dust emission. We apply a number of statistical tools to the data sets, including the probability density function of column density, the Velocity Gradient Technique (VGT), the polarization fraction analysis (PFA), and both the Davis–Chandrasekhar-Fermi (DCF) method and Differential Measure Analysis (DMA) technique. Our main discoveries are:
\begin{enumerate}
    \item The PDF of L1688's column density exhibits two distinguished power-law distributions suggesting different star formation activities (or rates) in the two corresponding density ranges. From the power-law PDF, we find L1688 is likely undergoing gravitational contraction at scale $\approx1.0$~pc and gravitational collapse at scale $\approx0.2$~pc.
    \item Using the VGT method, we found that the magnetic fields associated with H I foreground/background and molecular cloud L1688 are statistically coherent. 
    \item For the small Oph A clump, we found that VGT-C$^{18}$O measurement is perpendicular to the magnetic field derived from HAWC+ polarization. It suggests the presence of the gravitational collapse in Oph A, particularly at a volume density larger than $n_{\rm H_2}\ge10^4$ cm$^{-3}$.
    \item We used the PFA method to estimate the magnetic field's inclination angle distribution in L1688. We find the low-intensity tail has a mean inclination angle $\approx55^\circ$, while the central dense clump's mean inclination angle is $\approx30^\circ$.
    \item Based on the DCF method and the distribution of estimated inclination angles, we showed that L1688 has a relatively strong total magnetic field with a mean strength $\approx$~\SI{135}{\micro G}.
    \item We tested the new method, i.e., the Differential Measure Approach, for estimating magnetic field strength in L1688. The mean value of the total magnetic field is $\approx$~\SI{75}{\micro G}, smaller than that from the DCF's estimation.
    \item We found the strong magnetic field in L1688 regulates the large-scale gravitational contraction so that it mainly takes place along the magnetic field. Consequently, we observed the a power-law relation $B_{\rm tot}\propto n_{\rm H}^{1/2}$ when volume density is approximately less than $6.0\times10^3$ cm$^{-3}$.
    \item Based on the estimated total magnetic field strength, we found that gravitational contraction is present in a strongly magnetized and supersonic cloud.
\end{enumerate}

\section*{Acknowledgements}
Y.H. and A.L. acknowledge the support of NASA ATP AAH7546 and ALMA SOSPADA-016. Financial support for this work was provided by NASA through award 09\_0231 issued by the Universities Space Research Association, Inc. (USRA). This work used SDSC Expanse CPU at SDSC through allocations PHY230032, PHY230033, PHY230091, and PHY230105 from the Advanced Cyberinfrastructure Coordination Ecosystem: Services \& Support (ACCESS) program, which is supported by National Science Foundation grants \#2138259, \#2138286, \#2138307, \#2137603, and \#2138296. 
\section*{Data Availability}
The data underlying this article will be shared on reasonable request to the corresponding author. 


\bibliographystyle{mnras}
\bibliography{example} 

\begin{thebibliography}{}
\makeatletter
\relax
\def\mn@urlcharsother{\let\do\@makeother \do\$\do\&\do\#\do\^\do\_\do\%\do\~}
\def\mn@doi{\begingroup\mn@urlcharsother \@ifnextchar [ {\mn@doi@}
  {\mn@doi@[]}}
\def\mn@doi@[#1]#2{\def\@tempa{#1}\ifx\@tempa\@empty \href
  {http://dx.doi.org/#2} {doi:#2}\else \href {http://dx.doi.org/#2} {#1}\fi
  \endgroup}
\def\mn@eprint#1#2{\mn@eprint@#1:#2::\@nil}
\def\mn@eprint@arXiv#1{\href {http://arxiv.org/abs/#1} {{\tt arXiv:#1}}}
\def\mn@eprint@dblp#1{\href {http://dblp.uni-trier.de/rec/bibtex/#1.xml}
  {dblp:#1}}
\def\mn@eprint@#1:#2:#3:#4\@nil{\def\@tempa {#1}\def\@tempb {#2}\def\@tempc
  {#3}\ifx \@tempc \@empty \let \@tempc \@tempb \let \@tempb \@tempa \fi \ifx
  \@tempb \@empty \def\@tempb {arXiv}\fi \@ifundefined
  {mn@eprint@\@tempb}{\@tempb:\@tempc}{\expandafter \expandafter \csname
  mn@eprint@\@tempb\endcsname \expandafter{\@tempc}}}

\bibitem[\protect\citeauthoryear{{Alina} et~al.,}{{Alina}
  et~al.}{2022}]{Alina20}
{Alina} D.,  et~al., 2022, A\&A, \href
  {https://ui.adsabs.harvard.edu/abs/2020arXiv200715344A} {659, A90}

\bibitem[\protect\citeauthoryear{{Alves}, {Frau}, {Girart}, {Franco}, {Santos}
  \& {Wiesemeyer}}{{Alves} et~al.}{2014}]{2014A&A...569L...1A}
{Alves} F.~O.,  {Frau} P.,  {Girart} J.~M.,  {Franco} G.~A.~P.,  {Santos}
  F.~P.,   {Wiesemeyer} H.,  2014, \mn@doi [\aap]
  {10.1051/0004-6361/201424678}, \href
  {https://ui.adsabs.harvard.edu/abs/2014A&A...569L...1A} {569, L1}

\bibitem[\protect\citeauthoryear{{Andersson}, {Lazarian}  \&
  {Vaillancourt}}{{Andersson} et~al.}{2015}]{2015ARA&A..53..501A}
{Andersson} B.~G.,  {Lazarian} A.,   {Vaillancourt} J.~E.,  2015, ARAA, \href
  {https://ui.adsabs.harvard.edu/abs/2015ARA&A..53..501A} {53, 501}

\bibitem[\protect\citeauthoryear{{Andr{\'e}} et~al.,}{{Andr{\'e}}
  et~al.}{2010}]{2010AA...518L.102A}
{Andr{\'e}} P.,  et~al., 2010, \mn@doi [\aap] {10.1051/0004-6361/201014666},
  \href {https://ui.adsabs.harvard.edu/abs/2010A&A...518L.102A} {518, L102}

\bibitem[\protect\citeauthoryear{{Ballesteros-Paredes}, {V{\'a}zquez-Semadeni},
  {Gazol}, {Hartmann}, {Heitsch}  \& {Col{\'\i}n}}{{Ballesteros-Paredes}
  et~al.}{2011}]{2011MNRAS.416.1436B}
{Ballesteros-Paredes} J.,  {V{\'a}zquez-Semadeni} E.,  {Gazol} A.,  {Hartmann}
  L.~W.,  {Heitsch} F.,   {Col{\'\i}n} P.,  2011, \mn@doi [\mnras]
  {10.1111/j.1365-2966.2011.19141.x}, \href
  {https://ui.adsabs.harvard.edu/abs/2011MNRAS.416.1436B} {416, 1436}

\bibitem[\protect\citeauthoryear{{Bieging}, {Peters}  \& {Kang}}{{Bieging}
  et~al.}{2010}]{2010ApJS..191..232B}
{Bieging} J.~H.,  {Peters} W.~L.,   {Kang} M.,  2010, \mn@doi [\apjs]
  {10.1088/0067-0049/191/2/232}, \href
  {https://ui.adsabs.harvard.edu/abs/2010ApJS..191..232B} {191, 232}

\bibitem[\protect\citeauthoryear{{Burkhart}}{{Burkhart}}{2018}]{2018ApJ...863..118B}
{Burkhart} B.,  2018, \mn@doi [\apj] {10.3847/1538-4357/aad002}, \href
  {https://ui.adsabs.harvard.edu/abs/2018ApJ...863..118B} {863, 118}

\bibitem[\protect\citeauthoryear{{Chandrasekhar} \& {Fermi}}{{Chandrasekhar} \&
  {Fermi}}{1953}]{1953ApJ...118..113C}
{Chandrasekhar} S.,  {Fermi} E.,  1953, \mn@doi [\apj] {10.1086/145731}, \href
  {https://ui.adsabs.harvard.edu/abs/1953ApJ...118..113C} {118, 113}

\bibitem[\protect\citeauthoryear{{Chen}, {King}, {Li}, {Fissel}  \&
  {Mazzei}}{{Chen} et~al.}{2019}]{2019MNRAS.485.3499C}
{Chen} C.-Y.,  {King} P.~K.,  {Li} Z.-Y.,  {Fissel} L.~M.,   {Mazzei} R.~R.,
  2019, \mn@doi [\mnras] {10.1093/mnras/stz618}, \href
  {https://ui.adsabs.harvard.edu/abs/2019MNRAS.485.3499C} {485, 3499}

\bibitem[\protect\citeauthoryear{{Chen}, {Li}, {Mazzei}, {Park}, {Fissel},
  {Chen}, {Klein}  \& {Li}}{{Chen} et~al.}{2022}]{2022arXiv220509134C}
{Chen} C.-Y.,  {Li} Z.-Y.,  {Mazzei} R.~R.,  {Park} J.,  {Fissel} L.~M.,
  {Chen} M. C.~Y.,  {Klein} R.~I.,   {Li} P.~S.,  2022, arXiv e-prints, \href
  {https://ui.adsabs.harvard.edu/abs/2022arXiv220509134C} {p. arXiv:2205.09134}

\bibitem[\protect\citeauthoryear{{Cho} \& {Lazarian}}{{Cho} \&
  {Lazarian}}{2003}]{2003MNRAS.345..325C}
{Cho} J.,  {Lazarian} A.,  2003, \mn@doi [\mnras]
  {10.1046/j.1365-8711.2003.06941.x}, \href
  {https://ui.adsabs.harvard.edu/abs/2003MNRAS.345..325C} {345, 325}

\bibitem[\protect\citeauthoryear{{Choudhury} et~al.,}{{Choudhury}
  et~al.}{2021}]{2021A&A...648A.114C}
{Choudhury} S.,  et~al., 2021, \mn@doi [\aap] {10.1051/0004-6361/202039897},
  \href {https://ui.adsabs.harvard.edu/abs/2021A&A...648A.114C} {648, A114}

\bibitem[\protect\citeauthoryear{{Collins}, {Kritsuk}, {Padoan}, {Li}, {Xu},
  {Ustyugov}  \& {Norman}}{{Collins} et~al.}{2012}]{2012ApJ...750...13C}
{Collins} D.~C.,  {Kritsuk} A.~G.,  {Padoan} P.,  {Li} H.,  {Xu} H.,
  {Ustyugov} S.~D.,   {Norman} M.~L.,  2012, \mn@doi [\apj]
  {10.1088/0004-637X/750/1/13}, \href
  {https://ui.adsabs.harvard.edu/abs/2012ApJ...750...13C} {750, 13}

\bibitem[\protect\citeauthoryear{{Crutcher}}{{Crutcher}}{1999}]{1999ApJ...520..706C}
{Crutcher} R.~M.,  1999, \mn@doi [\apj] {10.1086/307483}, \href
  {https://ui.adsabs.harvard.edu/abs/1999ApJ...520..706C} {520, 706}

\bibitem[\protect\citeauthoryear{{Crutcher}}{{Crutcher}}{2004}]{Crutcher04}
{Crutcher} R.~M.,  2004, in {Uyaniker} B.,  {Reich} W.,   {Wielebinski} R.,
  eds, The Magnetized Interstellar Medium. pp 123--132

\bibitem[\protect\citeauthoryear{{Crutcher}}{{Crutcher}}{2012a}]{2012ARA&A..50...29C}
{Crutcher} R.~M.,  2012a, \mn@doi [\araa]
  {10.1146/annurev-astro-081811-125514}, \href
  {https://ui.adsabs.harvard.edu/abs/2012ARA&A..50...29C} {50, 29}

\bibitem[\protect\citeauthoryear{{Crutcher}}{{Crutcher}}{2012b}]{Crutcher12}
{Crutcher} R.~M.,  2012b, \mn@doi [\araa]
  {10.1146/annurev-astro-081811-125514}, \href
  {https://ui.adsabs.harvard.edu/abs/2012ARA&A..50...29C} {50, 29}

\bibitem[\protect\citeauthoryear{{Davis}}{{Davis}}{1951}]{1951PhRv...81..890D}
{Davis} L.,  1951, \mn@doi [Physical Review] {10.1103/PhysRev.81.890.2}, \href
  {https://ui.adsabs.harvard.edu/abs/1951PhRv...81..890D} {81, 890}

\bibitem[\protect\citeauthoryear{{Dunham} et~al.,}{{Dunham}
  et~al.}{2015}]{2015ApJS..220...11D}
{Dunham} M.~M.,  et~al., 2015, \mn@doi [\apjs] {10.1088/0067-0049/220/1/11},
  \href {https://ui.adsabs.harvard.edu/abs/2015ApJS..220...11D} {220, 11}

\bibitem[\protect\citeauthoryear{{Elmegreen}}{{Elmegreen}}{1993}]{1993ApJ...419L..29E}
{Elmegreen} B.~G.,  1993, \mn@doi [\apjl] {10.1086/187129}, \href
  {https://ui.adsabs.harvard.edu/abs/1993ApJ...419L..29E} {419, L29}

\bibitem[\protect\citeauthoryear{{Falceta-Gon{\c{c}}alves}, {Lazarian}  \&
  {Kowal}}{{Falceta-Gon{\c{c}}alves} et~al.}{2008}]{2008ApJ...679..537F}
{Falceta-Gon{\c{c}}alves} D.,  {Lazarian} A.,   {Kowal} G.,  2008, \mn@doi
  [\apj] {10.1086/587479}, \href
  {https://ui.adsabs.harvard.edu/abs/2008ApJ...679..537F} {679, 537}

\bibitem[\protect\citeauthoryear{{Federrath} \& {Klessen}}{{Federrath} \&
  {Klessen}}{2012}]{2012ApJ...761..156F}
{Federrath} C.,  {Klessen} R.~S.,  2012, \mn@doi [\apj]
  {10.1088/0004-637X/761/2/156}, \href
  {https://ui.adsabs.harvard.edu/abs/2012ApJ...761..156F} {761, 156}

\bibitem[\protect\citeauthoryear{{Federrath} \& {Klessen}}{{Federrath} \&
  {Klessen}}{2013}]{2013ApJ...763...51F}
{Federrath} C.,  {Klessen} R.~S.,  2013, \mn@doi [\apj]
  {10.1088/0004-637X/763/1/51}, \href
  {https://ui.adsabs.harvard.edu/abs/2013ApJ...763...51F} {763, 51}

\bibitem[\protect\citeauthoryear{{Federrath} et~al.,}{{Federrath}
  et~al.}{2016}]{2016ApJ...832..143F}
{Federrath} C.,  et~al., 2016, \mn@doi [\apj] {10.3847/0004-637X/832/2/143},
  \href {https://ui.adsabs.harvard.edu/abs/2016ApJ...832..143F} {832, 143}

\bibitem[\protect\citeauthoryear{{Fiedler} \& {Mouschovias}}{{Fiedler} \&
  {Mouschovias}}{1993}]{1993ApJ...415..680F}
{Fiedler} R.~A.,  {Mouschovias} T.~C.,  1993, \mn@doi [\apj] {10.1086/173193},
  \href {https://ui.adsabs.harvard.edu/abs/1993ApJ...415..680F} {415, 680}

\bibitem[\protect\citeauthoryear{Fisher}{Fisher}{1995}]{fisher1995statistical}
Fisher N.,  1995, Statistical Analysis of Circular Data.
Statistical Analysis of Circular Data, Cambridge University Press, \url
  {https://books.google.com/books?id=wGPj3EoFdJwC}

\bibitem[\protect\citeauthoryear{{Frau}, {Galli}  \& {Girart}}{{Frau}
  et~al.}{2011}]{2011A&A...535A..44F}
{Frau} P.,  {Galli} D.,   {Girart} J.~M.,  2011, \mn@doi [\aap]
  {10.1051/0004-6361/201117813}, \href
  {https://ui.adsabs.harvard.edu/abs/2011A&A...535A..44F} {535, A44}

\bibitem[\protect\citeauthoryear{{Galli} \& {Shu}}{{Galli} \&
  {Shu}}{1993}]{1993ApJ...417..243G}
{Galli} D.,  {Shu} F.~H.,  1993, \mn@doi [\apj] {10.1086/173306}, \href
  {https://ui.adsabs.harvard.edu/abs/1993ApJ...417..243G} {417, 243}

\bibitem[\protect\citeauthoryear{{Goldreich} \& {Sridhar}}{{Goldreich} \&
  {Sridhar}}{1995}]{GS95}
{Goldreich} P.,  {Sridhar} S.,  1995, \mn@doi [\apj] {10.1086/175121}, \href
  {https://ui.adsabs.harvard.edu/abs/1995ApJ...438..763G} {438, 763}

\bibitem[\protect\citeauthoryear{{Gonz{\'a}lez-Casanova} \&
  {Lazarian}}{{Gonz{\'a}lez-Casanova} \& {Lazarian}}{2017}]{GL17}
{Gonz{\'a}lez-Casanova} D.~F.,  {Lazarian} A.,  2017, \apj, \href
  {https://ui.adsabs.harvard.edu/abs/2017ApJ...835...41G} {835, 41}

\bibitem[\protect\citeauthoryear{{Goodman} \& {Heiles}}{{Goodman} \&
  {Heiles}}{1994}]{1994ApJ...424..208G}
{Goodman} A.~A.,  {Heiles} C.,  1994, \mn@doi [\apj] {10.1086/173884}, \href
  {https://ui.adsabs.harvard.edu/abs/1994ApJ...424..208G} {424, 208}

\bibitem[\protect\citeauthoryear{{G{\"u}ver} \& {{\"O}zel}}{{G{\"u}ver} \&
  {{\"O}zel}}{2009}]{2009MNRAS.400.2050G}
{G{\"u}ver} T.,  {{\"O}zel} F.,  2009, \mn@doi [\mnras]
  {10.1111/j.1365-2966.2009.15598.x}, \href
  {https://ui.adsabs.harvard.edu/abs/2009MNRAS.400.2050G} {400, 2050}

\bibitem[\protect\citeauthoryear{{Hartmann}, {Ballesteros-Paredes}  \&
  {Bergin}}{{Hartmann} et~al.}{2001}]{2001ApJ...562..852H}
{Hartmann} L.,  {Ballesteros-Paredes} J.,   {Bergin} E.~A.,  2001, \mn@doi
  [\apj] {10.1086/323863}, \href
  {https://ui.adsabs.harvard.edu/abs/2001ApJ...562..852H} {562, 852}

\bibitem[\protect\citeauthoryear{{Hildebrand}, {Kirby}, {Dotson}, {Houde}  \&
  {Vaillancourt}}{{Hildebrand} et~al.}{2009}]{2009ApJ...696..567H}
{Hildebrand} R.~H.,  {Kirby} L.,  {Dotson} J.~L.,  {Houde} M.,   {Vaillancourt}
  J.~E.,  2009, \mn@doi [\apj] {10.1088/0004-637X/696/1/567}, \href
  {https://ui.adsabs.harvard.edu/abs/2009ApJ...696..567H} {696, 567}

\bibitem[\protect\citeauthoryear{{Hoang}, {Tram}, {Lee}, {Diep}  \&
  {Ngoc}}{{Hoang} et~al.}{2021}]{2021ApJ...908..218H}
{Hoang} T.,  {Tram} L.~N.,  {Lee} H.,  {Diep} P.~N.,   {Ngoc} N.~B.,  2021,
  \mn@doi [\apj] {10.3847/1538-4357/abd54f}, \href
  {https://ui.adsabs.harvard.edu/abs/2021ApJ...908..218H} {908, 218}

\bibitem[\protect\citeauthoryear{{Hoang} et~al.,}{{Hoang}
  et~al.}{2022}]{2022ApJ...929...27H}
{Hoang} T.~D.,  et~al., 2022, \mn@doi [\apj] {10.3847/1538-4357/ac5abf}, \href
  {https://ui.adsabs.harvard.edu/abs/2022ApJ...929...27H} {929, 27}

\bibitem[\protect\citeauthoryear{{Houde}, {Vaillancourt}, {Hildebrand},
  {Chitsazzadeh}  \& {Kirby}}{{Houde} et~al.}{2009}]{2009ApJ...706.1504H}
{Houde} M.,  {Vaillancourt} J.~E.,  {Hildebrand} R.~H.,  {Chitsazzadeh} S.,
  {Kirby} L.,  2009, \mn@doi [\apj] {10.1088/0004-637X/706/2/1504}, \href
  {https://ui.adsabs.harvard.edu/abs/2009ApJ...706.1504H} {706, 1504}

\bibitem[\protect\citeauthoryear{{Hu} \& {Lazarian}}{{Hu} \&
  {Lazarian}}{2021}]{2021MNRAS.502.1768H}
{Hu} Y.,  {Lazarian} A.,  2021, \mn@doi [\mnras] {10.1093/mnras/stab087}, \href
  {https://ui.adsabs.harvard.edu/abs/2021MNRAS.502.1768H} {502, 1768}

\bibitem[\protect\citeauthoryear{{Hu} \& {Lazarian}}{{Hu} \&
  {Lazarian}}{2023}]{2022arXiv220309745H}
{Hu} Y.,  {Lazarian} A.,  2023, \mn@doi [\mnras] {10.1093/mnras/stac3744},
  \href {https://ui.adsabs.harvard.edu/abs/2023MNRAS.519.3736H} {519, 3736}

\bibitem[\protect\citeauthoryear{{Hu}, {Yuen}  \& {Lazarian}}{{Hu}
  et~al.}{2018}]{HYL18}
{Hu} Y.,  {Yuen} K.~H.,   {Lazarian} A.,  2018, \mn@doi [\mnras]
  {10.1093/mnras/sty1807}, \href
  {https://ui.adsabs.harvard.edu/abs/2018MNRAS.480.1333H} {480, 1333}

\bibitem[\protect\citeauthoryear{{Hu} et~al.,}{{Hu} et~al.}{2019}]{Hu19a}
{Hu} Y.,  et~al., 2019, \mn@doi [Nature Astronomy] {10.1038/s41550-019-0769-0},
  \href {https://ui.adsabs.harvard.edu/abs/2019NatAs...3..776H} {3, 776}

\bibitem[\protect\citeauthoryear{{Hu}, {Lazarian}  \& {Yuen}}{{Hu}
  et~al.}{2020a}]{HLY20}
{Hu} Y.,  {Lazarian} A.,   {Yuen} K.~H.,  2020a, \mn@doi [\apj]
  {10.3847/1538-4357/ab9948}, \href
  {https://ui.adsabs.harvard.edu/abs/2020ApJ...897..123H} {897, 123}

\bibitem[\protect\citeauthoryear{{Hu}, {Lazarian}  \& {Bialy}}{{Hu}
  et~al.}{2020b}]{2020ApJ...905..129H}
{Hu} Y.,  {Lazarian} A.,   {Bialy} S.,  2020b, \mn@doi [\apj]
  {10.3847/1538-4357/abc3c6}, \href
  {https://ui.adsabs.harvard.edu/abs/2020ApJ...905..129H} {905, 129}

\bibitem[\protect\citeauthoryear{{Hu}, {Xu}  \& {Lazarian}}{{Hu}
  et~al.}{2021a}]{2021ApJ...911...37H}
{Hu} Y.,  {Xu} S.,   {Lazarian} A.,  2021a, \mn@doi [\apj]
  {10.3847/1538-4357/abea18}, \href
  {https://ui.adsabs.harvard.edu/abs/2021ApJ...911...37H} {911, 37}

\bibitem[\protect\citeauthoryear{{Hu}, {Lazarian}  \& {Stanimirovi{\'c}}}{{Hu}
  et~al.}{2021b}]{HLS21}
{Hu} Y.,  {Lazarian} A.,   {Stanimirovi{\'c}} S.,  2021b, \mn@doi [\apj]
  {10.3847/1538-4357/abedb7}, \href
  {https://ui.adsabs.harvard.edu/abs/2021ApJ...912....2H} {912, 2}

\bibitem[\protect\citeauthoryear{{Hu}, {Lazarian}  \& {Wang}}{{Hu}
  et~al.}{2022}]{2022MNRAS.511..829H}
{Hu} Y.,  {Lazarian} A.,   {Wang} Q.~D.,  2022, \mnras, 511, 829

\bibitem[\protect\citeauthoryear{{Hwang} et~al.,}{{Hwang}
  et~al.}{2021}]{2021ApJ...913...85H}
{Hwang} J.,  et~al., 2021, \mn@doi [\apj] {10.3847/1538-4357/abf3c4}, \href
  {https://ui.adsabs.harvard.edu/abs/2021ApJ...913...85H} {913, 85}

\bibitem[\protect\citeauthoryear{{Jones}, {Bagley}, {Krejny}, {Andersson}  \&
  {Bastien}}{{Jones} et~al.}{2015}]{2015AJ....149...31J}
{Jones} T.~J.,  {Bagley} M.,  {Krejny} M.,  {Andersson} B.~G.,   {Bastien} P.,
  2015, \mn@doi [\aj] {10.1088/0004-6256/149/1/31}, \href
  {https://ui.adsabs.harvard.edu/abs/2015AJ....149...31J} {149, 31}

\bibitem[\protect\citeauthoryear{{Kandel}, {Lazarian}  \& {Pogosyan}}{{Kandel}
  et~al.}{2017}]{2017MNRAS.464.3617K}
{Kandel} D.,  {Lazarian} A.,   {Pogosyan} D.,  2017, \mn@doi [\mnras]
  {10.1093/mnras/stw2512}, \href
  {https://ui.adsabs.harvard.edu/abs/2017MNRAS.464.3617K} {464, 3617}

\bibitem[\protect\citeauthoryear{{Kauffmann}, {Bertoldi}, {Bourke}, {Evans}  \&
  {Lee}}{{Kauffmann} et~al.}{2008}]{2008A&A...487..993K}
{Kauffmann} J.,  {Bertoldi} F.,  {Bourke} T.~L.,  {Evans} N.~J. I.,   {Lee}
  C.~W.,  2008, \mn@doi [\aap] {10.1051/0004-6361:200809481}, \href
  {https://ui.adsabs.harvard.edu/abs/2008A&A...487..993K} {487, 993}

\bibitem[\protect\citeauthoryear{{Klessen}, {Heitsch}  \& {Mac Low}}{{Klessen}
  et~al.}{2000}]{2000ApJ...535..887K}
{Klessen} R.~S.,  {Heitsch} F.,   {Mac Low} M.-M.,  2000, \mn@doi [\apj]
  {10.1086/308891}, \href
  {https://ui.adsabs.harvard.edu/abs/2000ApJ...535..887K} {535, 887}

\bibitem[\protect\citeauthoryear{{K{\"o}rtgen}, {Federrath}  \&
  {Banerjee}}{{K{\"o}rtgen} et~al.}{2019}]{2019MNRAS.482.5233K}
{K{\"o}rtgen} B.,  {Federrath} C.,   {Banerjee} R.,  2019, \mn@doi [\mnras]
  {10.1093/mnras/sty3071}, \href
  {https://ui.adsabs.harvard.edu/abs/2019MNRAS.482.5233K} {482, 5233}

\bibitem[\protect\citeauthoryear{{Kwon} et~al.,}{{Kwon}
  et~al.}{2018}]{2018ApJ...859....4K}
{Kwon} J.,  et~al., 2018, \mn@doi [\apj] {10.3847/1538-4357/aabd82}, \href
  {https://ui.adsabs.harvard.edu/abs/2018ApJ...859....4K} {859, 4}

\bibitem[\protect\citeauthoryear{{Ladjelate} et~al.,}{{Ladjelate}
  et~al.}{2020}]{2020A&A...638A..74L}
{Ladjelate} B.,  et~al., 2020, \mn@doi [\aap] {10.1051/0004-6361/201936442},
  \href {https://ui.adsabs.harvard.edu/abs/2020A&A...638A..74L} {638, A74}

\bibitem[\protect\citeauthoryear{{Larson}}{{Larson}}{1981}]{1981MNRAS.194..809L}
{Larson} R.~B.,  1981, \mn@doi [\mnras] {10.1093/mnras/194.4.809}, \href
  {https://ui.adsabs.harvard.edu/abs/1981MNRAS.194..809L} {194, 809}

\bibitem[\protect\citeauthoryear{{Lazarian}}{{Lazarian}}{2005}]{2005AIPC..784...42L}
{Lazarian} A.,  2005, in {de Gouveia dal Pino} E.~M.,  {Lugones} G.,
  {Lazarian} A.,  eds,  American Institute of Physics Conference Series Vol.
  784, Magnetic Fields in the Universe: From Laboratory and Stars to Primordial
  Structures.. pp 42--53 (\mn@eprint {arXiv} {astro-ph/0505574}),
  \mn@doi{10.1063/1.2077170}

\bibitem[\protect\citeauthoryear{{Lazarian}}{{Lazarian}}{2014}]{2014SSRv..181....1L}
{Lazarian} A.,  2014, \mn@doi [\ssr] {10.1007/s11214-013-0031-5}, \href
  {https://ui.adsabs.harvard.edu/abs/2014SSRv..181....1L} {181, 1}

\bibitem[\protect\citeauthoryear{{Lazarian} \& {Hoang}}{{Lazarian} \&
  {Hoang}}{2007}]{2007MNRAS.378..910L}
{Lazarian} A.,  {Hoang} T.,  2007, \mn@doi [\mnras]
  {10.1111/j.1365-2966.2007.11817.x}, \href
  {https://ui.adsabs.harvard.edu/abs/2007MNRAS.378..910L} {378, 910}

\bibitem[\protect\citeauthoryear{{Lazarian} \& {Pogosyan}}{{Lazarian} \&
  {Pogosyan}}{2000}]{LP00}
{Lazarian} A.,  {Pogosyan} D.,  2000, ApJ, \href
  {https://ui.adsabs.harvard.edu/abs/2000ApJ...537..720L} {537, 720}

\bibitem[\protect\citeauthoryear{{Lazarian} \& {Vishniac}}{{Lazarian} \&
  {Vishniac}}{1999}]{LV99}
{Lazarian} A.,  {Vishniac} E.~T.,  1999, \mn@doi [\apj] {10.1086/307233}, \href
  {https://ui.adsabs.harvard.edu/abs/1999ApJ...517..700L} {517, 700}

\bibitem[\protect\citeauthoryear{{Lazarian} \& {Yuen}}{{Lazarian} \&
  {Yuen}}{2018}]{LY18a}
{Lazarian} A.,  {Yuen} K.~H.,  2018, \mn@doi [\apj] {10.3847/1538-4357/aaa241},
  \href {https://ui.adsabs.harvard.edu/abs/2018ApJ...853...96L} {853, 96}

\bibitem[\protect\citeauthoryear{{Lazarian}, {Esquivel}  \&
  {Crutcher}}{{Lazarian} et~al.}{2012}]{2012ApJ...757..154L}
{Lazarian} A.,  {Esquivel} A.,   {Crutcher} R.,  2012, \mn@doi [\apj]
  {10.1088/0004-637X/757/2/154}, \href
  {https://ui.adsabs.harvard.edu/abs/2012ApJ...757..154L} {757, 154}

\bibitem[\protect\citeauthoryear{{Lazarian}, {Yuen}, {Ho}, {Chen}, {Lazarian},
  {Lu}, {Yang}  \& {Hu}}{{Lazarian} et~al.}{2018}]{Lazarian18}
{Lazarian} A.,  {Yuen} K.~H.,  {Ho} K.~W.,  {Chen} J.,  {Lazarian} V.,  {Lu}
  Z.,  {Yang} B.,   {Hu} Y.,  2018, \mn@doi [\apj] {10.3847/1538-4357/aad7ff},
  \href {https://ui.adsabs.harvard.edu/abs/2018ApJ...865...46L} {865, 46}

\bibitem[\protect\citeauthoryear{{Lazarian}, {Yuen}  \& {Pogosyan}}{{Lazarian}
  et~al.}{2022}]{2022arXiv220409731L}
{Lazarian} A.,  {Yuen} K.~H.,   {Pogosyan} D.,  2022, arXiv e-prints, \href
  {https://ui.adsabs.harvard.edu/abs/2022arXiv220409731L} {p. arXiv:2204.09731}

\bibitem[\protect\citeauthoryear{{Leroy} et~al.,}{{Leroy}
  et~al.}{2015}]{2015ApJ...801...25L}
{Leroy} A.~K.,  et~al., 2015, \mn@doi [\apj] {10.1088/0004-637X/801/1/25},
  \href {https://ui.adsabs.harvard.edu/abs/2015ApJ...801...25L} {801, 25}

\bibitem[\protect\citeauthoryear{{Li} \& {Goldsmith}}{{Li} \&
  {Goldsmith}}{2003}]{2003ApJ...585..823L}
{Li} D.,  {Goldsmith} P.~F.,  2003, \mn@doi [\apj] {10.1086/346227}, \href
  {https://ui.adsabs.harvard.edu/abs/2003ApJ...585..823L} {585, 823}

\bibitem[\protect\citeauthoryear{{Li}, {Lopez-Rodriguez}, {Ajeddig},
  {Andr{\'e}}, {McKee}, {Rho}  \& {Klein}}{{Li}
  et~al.}{2021}]{2021MNRAS.tmp.3119L}
{Li} P.~S.,  {Lopez-Rodriguez} E.,  {Ajeddig} H.,  {Andr{\'e}} P.,  {McKee}
  C.~F.,  {Rho} J.,   {Klein} R.~I.,  2021, \mn@doi [\mnras]
  {10.1093/mnras/stab3448}, \href
  {https://ui.adsabs.harvard.edu/abs/2021MNRAS.tmp.3119L} {}

\bibitem[\protect\citeauthoryear{{Liseau}, {Larsson}, {Bergman}, {Pagani},
  {Black}, {Hjalmarson}  \& {Justtanont}}{{Liseau}
  et~al.}{2010}]{2010AA...510A..98L}
{Liseau} R.,  {Larsson} B.,  {Bergman} P.,  {Pagani} L.,  {Black} J.~H.,
  {Hjalmarson} {\r{A}}.,   {Justtanont} K.,  2010, \mn@doi [\aap]
  {10.1051/0004-6361/200913567}, \href
  {https://ui.adsabs.harvard.edu/abs/2010A&A...510A..98L} {510, A98}

\bibitem[\protect\citeauthoryear{{Liu} et~al.,}{{Liu}
  et~al.}{2019}]{2019ApJ...877...43L}
{Liu} J.,  et~al., 2019, \mn@doi [\apj] {10.3847/1538-4357/ab0958}, \href
  {https://ui.adsabs.harvard.edu/abs/2019ApJ...877...43L} {877, 43}

\bibitem[\protect\citeauthoryear{{Liu}, {Qiu}  \& {Zhang}}{{Liu}
  et~al.}{2022}]{2022ApJ...925...30L}
{Liu} J.,  {Qiu} K.,   {Zhang} Q.,  2022, \mn@doi [\apj]
  {10.3847/1538-4357/ac3911}, \href
  {https://ui.adsabs.harvard.edu/abs/2022ApJ...925...30L} {925, 30}

\bibitem[\protect\citeauthoryear{{Mac Low} \& {Klessen}}{{Mac Low} \&
  {Klessen}}{2004}]{MK04}
{Mac Low} M.-M.,  {Klessen} R.~S.,  2004, \mn@doi [Reviews of Modern Physics]
  {10.1103/RevModPhys.76.125}, \href
  {https://ui.adsabs.harvard.edu/abs/2004RvMP...76..125M} {76, 125}

\bibitem[\protect\citeauthoryear{{McKee} \& {Ostriker}}{{McKee} \&
  {Ostriker}}{2007}]{MO07}
{McKee} C.~F.,  {Ostriker} E.~C.,  2007, \mn@doi [\araa]
  {10.1146/annurev.astro.45.051806.110602}, \href
  {https://ui.adsabs.harvard.edu/abs/2007ARA&A..45..565M} {45, 565}

\bibitem[\protect\citeauthoryear{{Mestel}}{{Mestel}}{1965}]{1965QJRAS...6..265M}
{Mestel} L.,  1965, \qjras, \href
  {https://ui.adsabs.harvard.edu/abs/1965QJRAS...6..265M} {6, 265}

\bibitem[\protect\citeauthoryear{{Mouschovias}}{{Mouschovias}}{1976}]{1976ApJ...207..141M}
{Mouschovias} T.~C.,  1976, \mn@doi [\apj] {10.1086/154478}, \href
  {https://ui.adsabs.harvard.edu/abs/1976ApJ...207..141M} {207, 141}

\bibitem[\protect\citeauthoryear{{Ortiz-Le{\'o}n} et~al.,}{{Ortiz-Le{\'o}n}
  et~al.}{2018}]{2018ApJ...869L..33O}
{Ortiz-Le{\'o}n} G.~N.,  et~al., 2018, \mn@doi [\apjl]
  {10.3847/2041-8213/aaf6ad}, \href
  {https://ui.adsabs.harvard.edu/abs/2018ApJ...869L..33O} {869, L33}

\bibitem[\protect\citeauthoryear{{Ostriker}, {Stone}  \& {Gammie}}{{Ostriker}
  et~al.}{2001}]{2001ApJ...546..980O}
{Ostriker} E.~C.,  {Stone} J.~M.,   {Gammie} C.~F.,  2001, \mn@doi [\apj]
  {10.1086/318290}, \href
  {https://ui.adsabs.harvard.edu/abs/2001ApJ...546..980O} {546, 980}

\bibitem[\protect\citeauthoryear{{Padoan}}{{Padoan}}{1995}]{1995MNRAS.277..377P}
{Padoan} P.,  1995, \mn@doi [\mnras] {10.1093/mnras/277.2.377}, \href
  {https://ui.adsabs.harvard.edu/abs/1995MNRAS.277..377P} {277, 377}

\bibitem[\protect\citeauthoryear{{Pattle} et~al.,}{{Pattle}
  et~al.}{2015}]{2015MNRAS.450.1094P}
{Pattle} K.,  et~al., 2015, \mn@doi [\mnras] {10.1093/mnras/stv376}, \href
  {https://ui.adsabs.harvard.edu/abs/2015MNRAS.450.1094P} {450, 1094}

\bibitem[\protect\citeauthoryear{{Pattle} et~al.,}{{Pattle}
  et~al.}{2019}]{2019ApJ...880...27P}
{Pattle} K.,  et~al., 2019, \mn@doi [\apj] {10.3847/1538-4357/ab286f}, \href
  {https://ui.adsabs.harvard.edu/abs/2019ApJ...880...27P} {880, 27}

\bibitem[\protect\citeauthoryear{{Pattle} et~al.,}{{Pattle}
  et~al.}{2021}]{2021ApJ...907...88P}
{Pattle} K.,  et~al., 2021, \mn@doi [\apj] {10.3847/1538-4357/abcc6c}, \href
  {https://ui.adsabs.harvard.edu/abs/2021ApJ...907...88P} {907, 88}

\bibitem[\protect\citeauthoryear{{Planck Collaboration} et~al.,}{{Planck
  Collaboration} et~al.}{2015a}]{2015A&A...576A.104P}
{Planck Collaboration} et~al., 2015a, \mn@doi [\aap]
  {10.1051/0004-6361/201424082}, \href
  {https://ui.adsabs.harvard.edu/abs/2015A&A...576A.104P} {576, A104}

\bibitem[\protect\citeauthoryear{{Planck Collaboration} et~al.,}{{Planck
  Collaboration} et~al.}{2015b}]{2015A&A...576A.105P}
{Planck Collaboration} et~al., 2015b, \mn@doi [\aap]
  {10.1051/0004-6361/201424086}, \href
  {https://ui.adsabs.harvard.edu/abs/2015A&A...576A.105P} {576, A105}

\bibitem[\protect\citeauthoryear{{Planck Collaboration} et~al.,}{{Planck
  Collaboration} et~al.}{2020a}]{2020A&A...641A...3P}
{Planck Collaboration} et~al., 2020a, \mn@doi [\aap]
  {10.1051/0004-6361/201832909}, \href
  {https://ui.adsabs.harvard.edu/abs/2020A&A...641A...3P} {641, A3}

\bibitem[\protect\citeauthoryear{{Planck Collaboration} et~al.,}{{Planck
  Collaboration} et~al.}{2020b}]{2020AA...641A..11P}
{Planck Collaboration} et~al., 2020b, \mn@doi [\aap]
  {10.1051/0004-6361/201832618}, \href
  {https://ui.adsabs.harvard.edu/abs/2020A&A...641A..11P} {641, A11}

\bibitem[\protect\citeauthoryear{{Planck Collaboration} et~al.,}{{Planck
  Collaboration} et~al.}{2020c}]{2020A&A...641A..12P}
{Planck Collaboration} et~al., 2020c, \mn@doi [\aap]
  {10.1051/0004-6361/201833885}, \href
  {https://ui.adsabs.harvard.edu/abs/2020A&A...641A..12P} {641, A12}

\bibitem[\protect\citeauthoryear{{Ridge} et~al.,}{{Ridge}
  et~al.}{2006}]{2006AJ....131.2921R}
{Ridge} N.~A.,  et~al., 2006, \mn@doi [\aj] {10.1086/503704}, \href
  {https://ui.adsabs.harvard.edu/abs/2006AJ....131.2921R} {131, 2921}

\bibitem[\protect\citeauthoryear{{Santos} et~al.,}{{Santos}
  et~al.}{2019}]{2019ApJ...882..113S}
{Santos} F.~P.,  et~al., 2019, \mn@doi [\apj] {10.3847/1538-4357/ab3407}, \href
  {https://ui.adsabs.harvard.edu/abs/2019ApJ...882..113S} {882, 113}

\bibitem[\protect\citeauthoryear{{Sato}, {Tamura}, {Nagata}, {Kaifu}, {Hough},
  {McLean}, {Garden}  \& {Gatley}}{{Sato} et~al.}{1988}]{1988MNRAS.230..321S}
{Sato} S.,  {Tamura} M.,  {Nagata} T.,  {Kaifu} N.,  {Hough} J.,  {McLean}
  I.~S.,  {Garden} R.~P.,   {Gatley} I.,  1988, \mn@doi [\mnras]
  {10.1093/mnras/230.2321}, \href
  {https://ui.adsabs.harvard.edu/abs/1988MNRAS.230..321S} {230, 321}

\bibitem[\protect\citeauthoryear{{Scott} \& {Black}}{{Scott} \&
  {Black}}{1980}]{1980ApJ...239..166S}
{Scott} E.~H.,  {Black} D.~C.,  1980, \mn@doi [\apj] {10.1086/158098}, \href
  {https://ui.adsabs.harvard.edu/abs/1980ApJ...239..166S} {239, 166}

\bibitem[\protect\citeauthoryear{{Seifried}, {Walch}, {Reissl}  \&
  {Ib{\'a}{\~n}ez-Mej{\'\i}a}}{{Seifried} et~al.}{2019}]{2019MNRAS.482.2697S}
{Seifried} D.,  {Walch} S.,  {Reissl} S.,   {Ib{\'a}{\~n}ez-Mej{\'\i}a} J.~C.,
  2019, \mn@doi [\mnras] {10.1093/mnras/sty2831}, \href
  {https://ui.adsabs.harvard.edu/abs/2019MNRAS.482.2697S} {482, 2697}

\bibitem[\protect\citeauthoryear{{Shimajiri} et~al.,}{{Shimajiri}
  et~al.}{2011}]{2011PASJ...63..105S}
{Shimajiri} Y.,  et~al., 2011, \mn@doi [\pasj] {10.1093/pasj/63.1.105}, \href
  {https://ui.adsabs.harvard.edu/abs/2011PASJ...63..105S} {63, 105}

\bibitem[\protect\citeauthoryear{{Skalidis}, {Sternberg}, {Beattie}, {Pavlidou}
   \& {Tassis}}{{Skalidis} et~al.}{2021}]{2021A&A...656A.118S}
{Skalidis} R.,  {Sternberg} J.,  {Beattie} J.~R.,  {Pavlidou} V.,   {Tassis}
  K.,  2021, \mn@doi [\aap] {10.1051/0004-6361/202142045}, \href
  {https://ui.adsabs.harvard.edu/abs/2021A&A...656A.118S} {656, A118}

\bibitem[\protect\citeauthoryear{{Soam} et~al.,}{{Soam}
  et~al.}{2018}]{2018ApJ...861...65S}
{Soam} A.,  et~al., 2018, \mn@doi [\apj] {10.3847/1538-4357/aac4a6}, \href
  {https://ui.adsabs.harvard.edu/abs/2018ApJ...861...65S} {861, 65}

\bibitem[\protect\citeauthoryear{{Sullivan}, {Fissel}, {King}, {Chen}, {Li}  \&
  {Soler}}{{Sullivan} et~al.}{2021}]{2021MNRAS.503.5006S}
{Sullivan} C.~H.,  {Fissel} L.~M.,  {King} P.~K.,  {Chen} C.~Y.,  {Li} Z.~Y.,
  {Soler} J.~D.,  2021, \mn@doi [\mnras] {10.1093/mnras/stab596}, \href
  {https://ui.adsabs.harvard.edu/abs/2021MNRAS.503.5006S} {503, 5006}

\bibitem[\protect\citeauthoryear{{Tahani}, {Plume}, {Brown}  \&
  {Kainulainen}}{{Tahani} et~al.}{2018}]{2018A&A...614A.100T}
{Tahani} M.,  {Plume} R.,  {Brown} J.~C.,   {Kainulainen} J.,  2018, \mn@doi
  [\aap] {10.1051/0004-6361/201732219}, \href
  {https://ui.adsabs.harvard.edu/abs/2018A&A...614A.100T} {614, A100}

\bibitem[\protect\citeauthoryear{{Tram} et~al.,}{{Tram}
  et~al.}{2022}]{2022arXiv220512084T}
{Tram} L.~N.,  et~al., 2022, arXiv e-prints, \href
  {https://ui.adsabs.harvard.edu/abs/2022arXiv220512084T} {p. arXiv:2205.12084}

\bibitem[\protect\citeauthoryear{{Troland}, {Crutcher}, {Goodman}, {Heiles},
  {Kazes}  \& {Myers}}{{Troland} et~al.}{1996}]{1996ApJ...471..302T}
{Troland} T.~H.,  {Crutcher} R.~M.,  {Goodman} A.~A.,  {Heiles} C.,  {Kazes}
  I.,   {Myers} P.~C.,  1996, \mn@doi [\apj] {10.1086/177970}, \href
  {https://ui.adsabs.harvard.edu/abs/1996ApJ...471..302T} {471, 302}

\bibitem[\protect\citeauthoryear{{V{\'a}zquez-Semadeni}, {Banerjee},
  {G{\'o}mez}, {Hennebelle}, {Duffin}  \& {Klessen}}{{V{\'a}zquez-Semadeni}
  et~al.}{2011}]{2011MNRAS.414.2511V}
{V{\'a}zquez-Semadeni} E.,  {Banerjee} R.,  {G{\'o}mez} G.~C.,  {Hennebelle}
  P.,  {Duffin} D.,   {Klessen} R.~S.,  2011, \mn@doi [\mnras]
  {10.1111/j.1365-2966.2011.18569.x}, \href
  {https://ui.adsabs.harvard.edu/abs/2011MNRAS.414.2511V} {414, 2511}

\bibitem[\protect\citeauthoryear{{V{\'a}zquez-Semadeni}, {Palau},
  {Ballesteros-Paredes}, {G{\'o}mez}  \&
  {Zamora-Avil{\'e}s}}{{V{\'a}zquez-Semadeni}
  et~al.}{2019}]{2019MNRAS.490.3061V}
{V{\'a}zquez-Semadeni} E.,  {Palau} A.,  {Ballesteros-Paredes} J.,  {G{\'o}mez}
  G.~C.,   {Zamora-Avil{\'e}s} M.,  2019, \mn@doi [\mnras]
  {10.1093/mnras/stz2736}, \href
  {https://ui.adsabs.harvard.edu/abs/2019MNRAS.490.3061V} {490, 3061}

\bibitem[\protect\citeauthoryear{{Vrba}}{{Vrba}}{1977}]{1977AJ.....82..198V}
{Vrba} F.~J.,  1977, \mn@doi [\aj] {10.1086/112031}, \href
  {https://ui.adsabs.harvard.edu/abs/1977AJ.....82..198V} {82, 198}

\bibitem[\protect\citeauthoryear{{Wardle} \& {Kronberg}}{{Wardle} \&
  {Kronberg}}{1974}]{1974ApJ...194..249W}
{Wardle} J.~F.~C.,  {Kronberg} P.~P.,  1974, \mn@doi [\apj] {10.1086/153240},
  \href {https://ui.adsabs.harvard.edu/abs/1974ApJ...194..249W} {194, 249}

\bibitem[\protect\citeauthoryear{{White} et~al.,}{{White}
  et~al.}{2015}]{2015MNRAS.447.1996W}
{White} G.~J.,  et~al., 2015, \mn@doi [\mnras] {10.1093/mnras/stu2323}, \href
  {https://ui.adsabs.harvard.edu/abs/2015MNRAS.447.1996W} {447, 1996}

\bibitem[\protect\citeauthoryear{{Whittet}, {Hough}, {Lazarian}  \&
  {Hoang}}{{Whittet} et~al.}{2008}]{2008ApJ...674..304W}
{Whittet} D.~C.~B.,  {Hough} J.~H.,  {Lazarian} A.,   {Hoang} T.,  2008,
  \mn@doi [\apj] {10.1086/525040}, \href
  {https://ui.adsabs.harvard.edu/abs/2008ApJ...674..304W} {674, 304}

\bibitem[\protect\citeauthoryear{{Wilking}, {Lebofsky}, {Rieke}  \&
  {Kemp}}{{Wilking} et~al.}{1979}]{1979AJ.....84..199W}
{Wilking} B.~A.,  {Lebofsky} M.~J.,  {Rieke} G.~H.,   {Kemp} J.~C.,  1979,
  \mn@doi [\aj] {10.1086/112408}, \href
  {https://ui.adsabs.harvard.edu/abs/1979AJ.....84..199W} {84, 199}

\bibitem[\protect\citeauthoryear{{Wilking}, {Gagn{\'e}}  \& {Allen}}{{Wilking}
  et~al.}{2008}]{2008hsf2.book..351W}
{Wilking} B.~A.,  {Gagn{\'e}} M.,   {Allen} L.~E.,  2008, in {Reipurth} B.,
  ed., , Vol.~5, Handbook of Star Forming Regions, Volume II.
p.~351

\bibitem[\protect\citeauthoryear{{Xu} \& {Hu}}{{Xu} \&
  {Hu}}{2021}]{2021ApJ...910...88X}
{Xu} S.,  {Hu} Y.,  2021, \mn@doi [\apj] {10.3847/1538-4357/abe403}, \href
  {https://ui.adsabs.harvard.edu/abs/2021ApJ...910...88X} {910, 88}

\bibitem[\protect\citeauthoryear{{Yuen} \& {Lazarian}}{{Yuen} \&
  {Lazarian}}{2017}]{YL17a}
{Yuen} K.~H.,  {Lazarian} A.,  2017, \mn@doi [\apjl]
  {10.3847/2041-8213/aa6255}, \href
  {https://ui.adsabs.harvard.edu/abs/2017ApJ...837L..24Y} {837, L24}

\bibitem[\protect\citeauthoryear{{Zucker} et~al.,}{{Zucker}
  et~al.}{2021}]{2021ApJ...919...35Z}
{Zucker} C.,  et~al., 2021, \mn@doi [\apj] {10.3847/1538-4357/ac1f96}, \href
  {https://ui.adsabs.harvard.edu/abs/2021ApJ...919...35Z} {919, 35}

\makeatother
\end{thebibliography}



\newpage
\appendix
\section{Uncertainty of $p$}
\begin{figure}
	\includegraphics[width=1.0\linewidth]{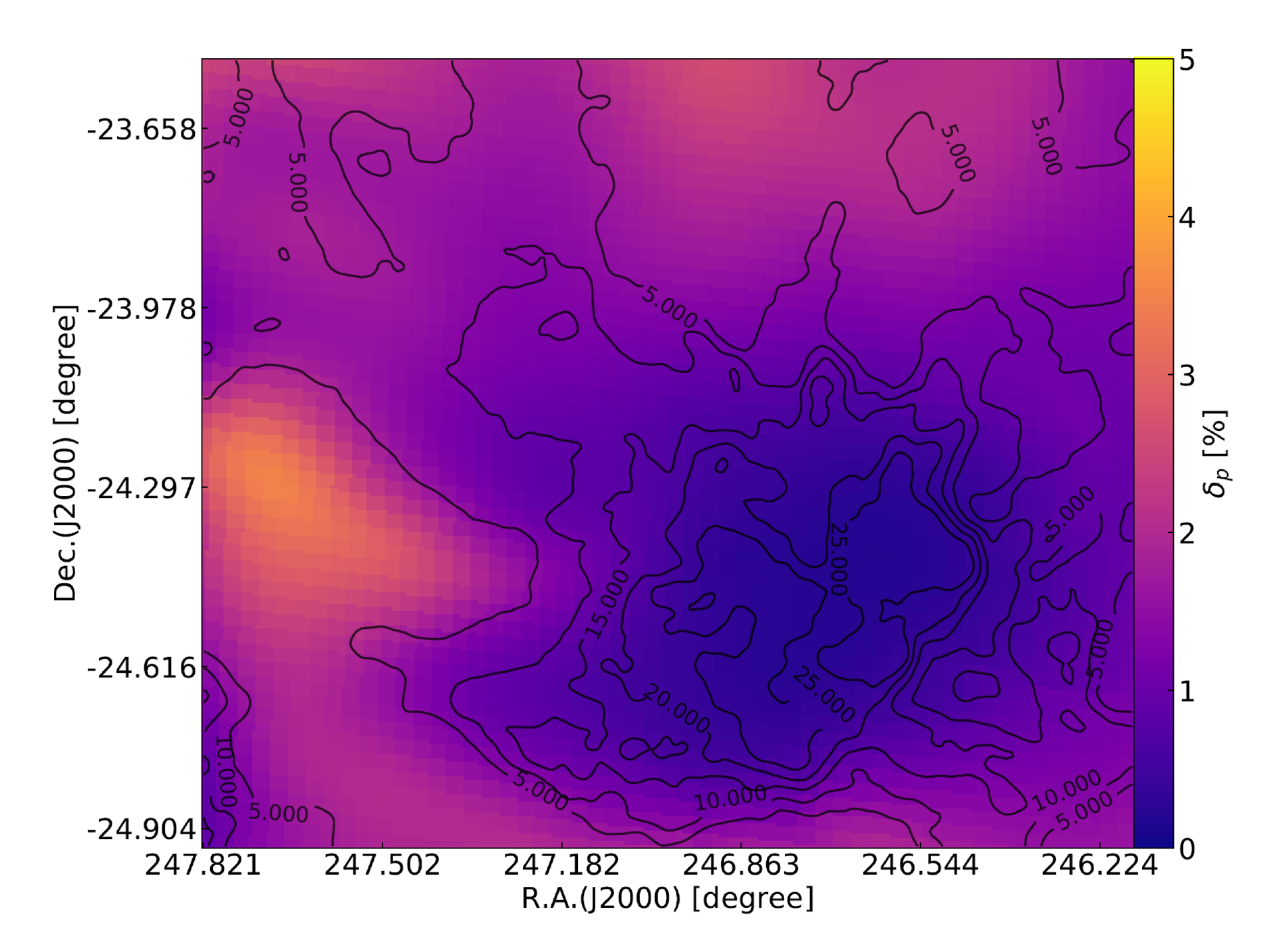}
    \caption{The uncertainty map of the polarization fraction $p$. Contours outline the intensity structures of $^{12}$CO starting from 5 K km/s}
    \label{fig:sigma_p}
\end{figure}

In Fig.~\ref{fig:sigma_p}, we present the uncertainty map of the polarization fraction $p$. The uncertainty is calculated from:
\begin{equation}
    \delta_p=(\frac{Q^2\delta Q^2+U^2\delta U^2}{I^2(Q^2+U^2)}+\frac{\delta I^2(Q^2+U^2)}{I^4})^{1/2},
\end{equation}
where $\delta I$, $\delta Q$, and $\delta U$ are the uncertainties in Stokes $I$, $Q$, $U$, respectively. Within the contours, the uncertainty is generally less than 1\%. Particularly, for the dense Oph A, B, C, E, F regions, the uncertainty is around 0.2\%.

\begin{figure}
	\includegraphics[width=1.0\linewidth]{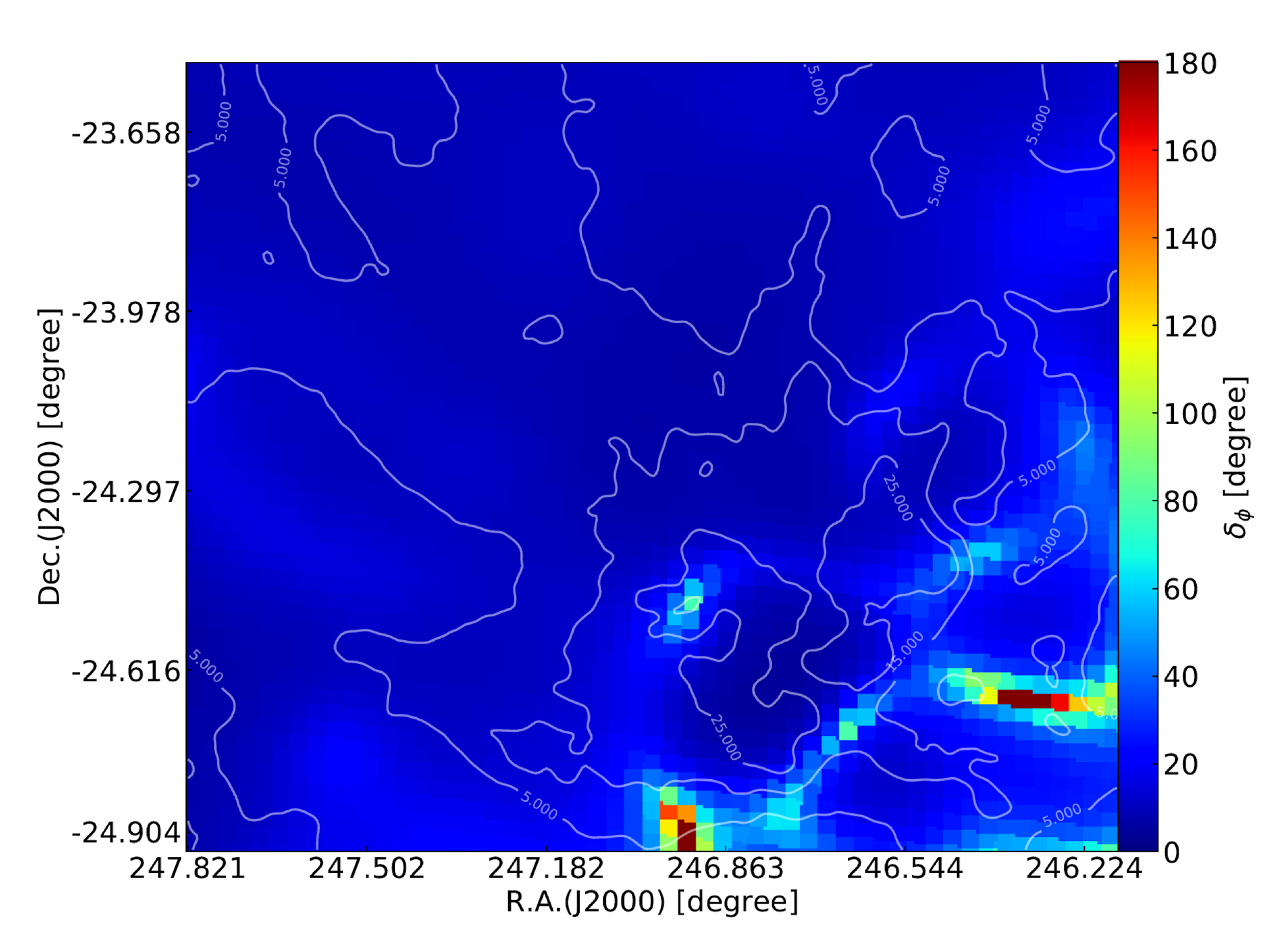}
    \caption{The uncertainty map of the polarization angle $\phi$. Contours outline the intensity structures of $^{12}$CO starting from 5 K km/s}
    \label{fig:sigma_phi}
\end{figure}

The uncertainty of polarization angle $\phi$ is:
\begin{equation}
    \delta_\phi=\frac{1}{2}\sqrt{\frac{Q^2\delta U^2+U^2\delta Q^2}{(Q^2+U^2)^2}}.
\end{equation}
We present the map of $\delta_\phi$ in Fig.~\ref{fig:sigma_phi}. The median value of $\delta_\phi$ is around $14^\circ$.

\section{Correlation of inclination angle and Planck emission intensity}
\begin{figure}
	\includegraphics[width=1.0\linewidth]{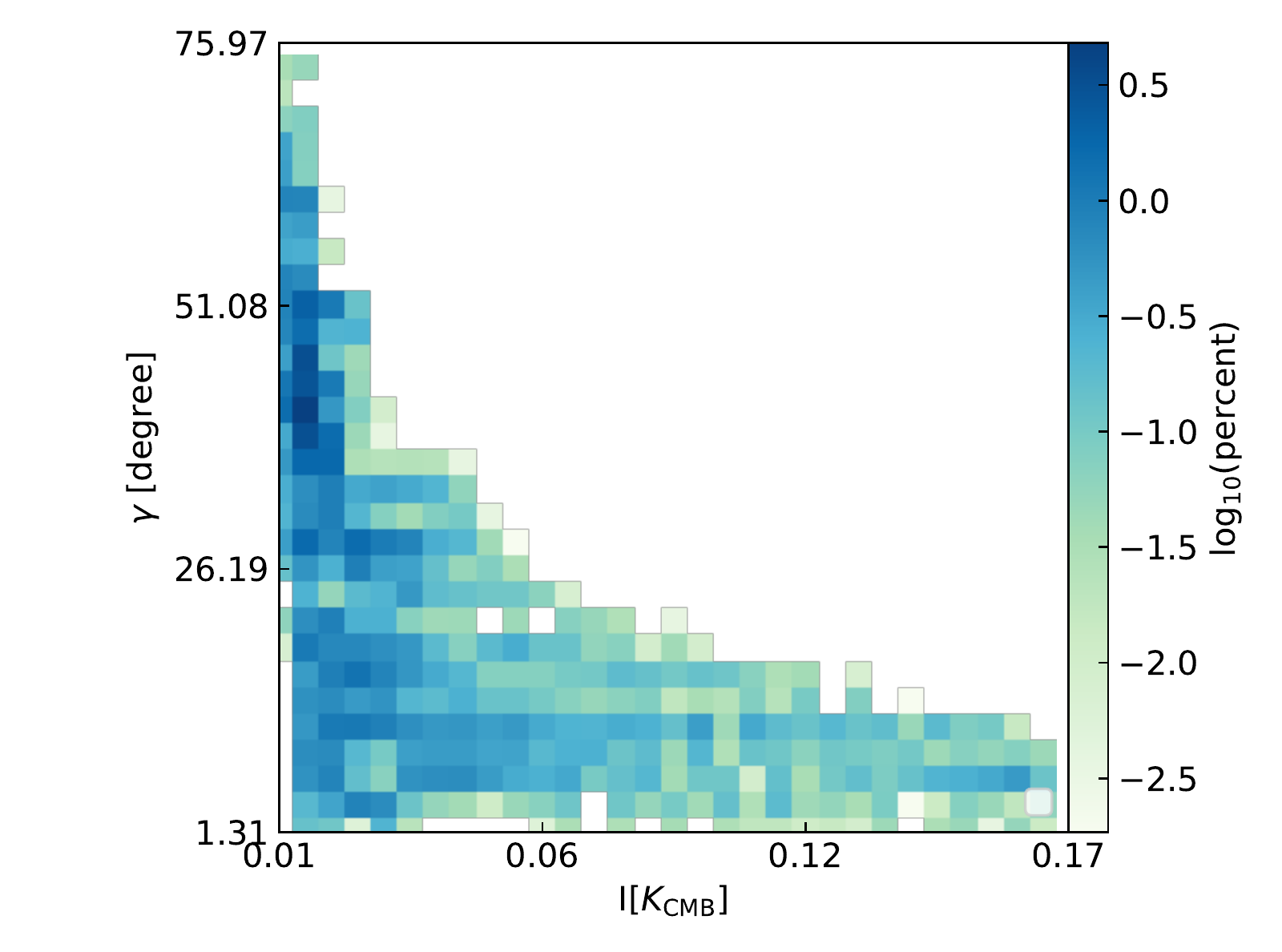}
    \caption{The 2D histogram of inclination angle derived from PFA and Planck emission intensity at 353 GHz. }
    \label{fig:Ivsgamma}
\end{figure}

Fig.~\ref{fig:Ivsgamma} presents the 2D histogram of inclination angle derived from PFA and Planck emission intensity. We see the inclination angle tends to be small values at large intensity. As we discussed in \S~\ref{sec:dis}, there are two possible reasons: (1) the magnetic fields has significant variations; or  (2) the loss of grain alignment at high extinction.

\begin{figure}
	\includegraphics[width=1.0\linewidth]{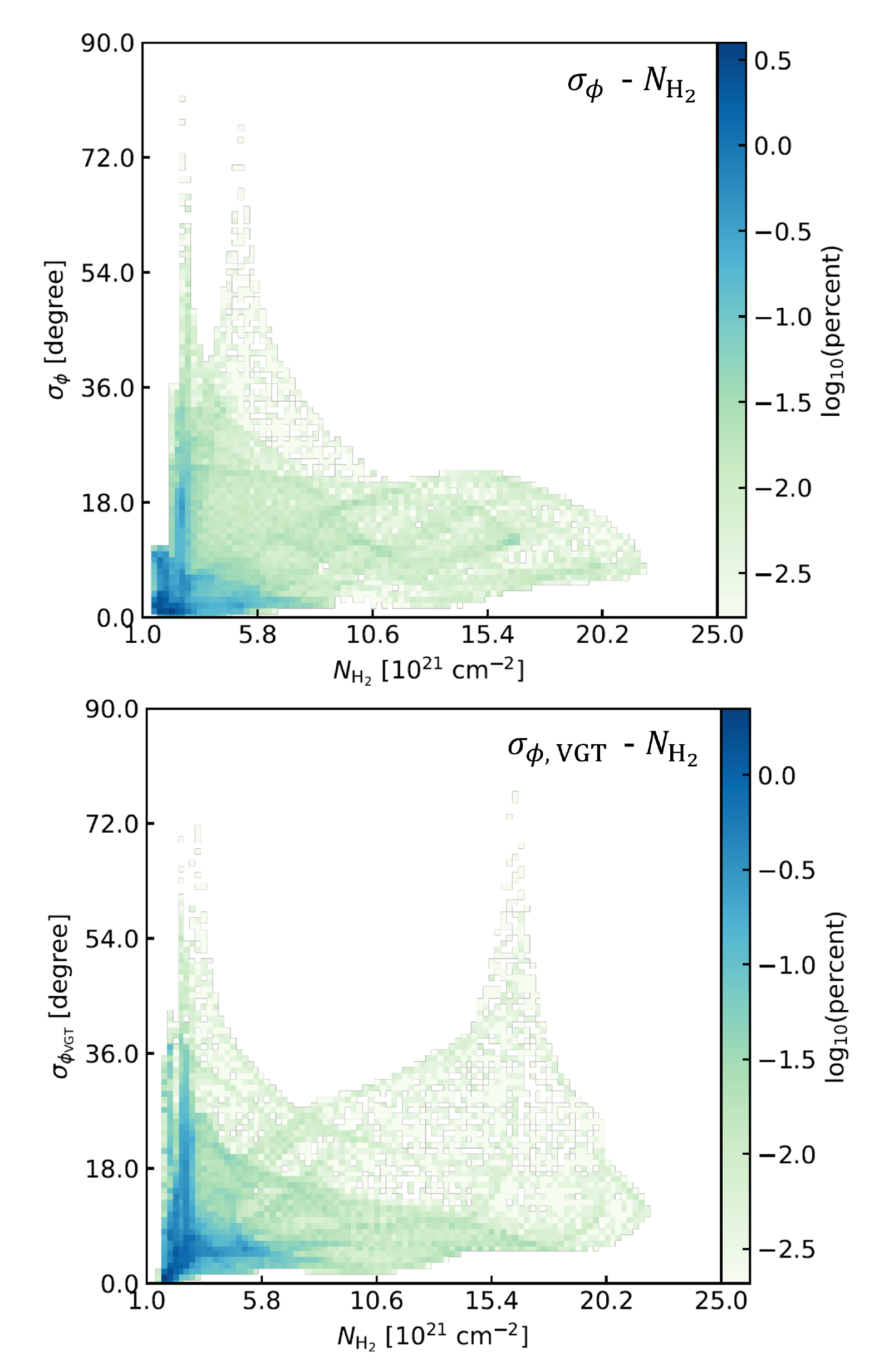}
    \caption{The 2D histogram of polarization angle dispersion (top), as well as VGT-$^{12}$CO angle dispersion (bottom), and column density.}
    \label{fig:pdisNH2}
\end{figure}

In Fig.~\ref{fig:pdisNH2}, we plot the 2D histogram of polarization angle dispersion and column density. We find in low-density regions ($N_{\rm H_2}<5.8\times10^{21}$~cm$^{-2}$), the dispersion of polarization angle is significant, corresponding to the small inclination angles observed in L1688's southwest (see Fig.~\ref{fig:gamma}). Similarly, we also plot the 2D histogram of VGT-$^{12}$CO angle dispersion, which is insensitive to grain alignment, and column density. Similar large angle dispersion associated with low-density is also observed. We, therefore, expect the depolarization due to field variation dominates L1688's southwest so the (effective) inclination angle is indeed small. 

In addition, we find the angle dispersion of VGT-$^{12}$CO is also significant at $N_{\rm H_2}\approx1.54\times10^{22}$~cm$^{-2}$. This corresponds to the regions of dense Oph A, B, and C cores and we expect self-gravity is responsible for the variation of VGT-$^{12}$CO.

\bsp	
\label{lastpage}
\end{document}